\documentclass[12pt,a4paper,dvips]{article}
\usepackage{a4p}
\usepackage{epsfig}
\usepackage{graphicx,rotating}
\usepackage{physics }
\usepackage{l3_titlePH,ifthen}
\usepackage{Lep}
\usepackage{cite}
\journalname{Phys. Lett. B}
\preprint{2005-004} 
\date{February 15, 2005}

\def\Q2 {Q^2}
\def\AP2 {P^2}
\def\q2 { q^2 }
\def\ap2 { p^2 }

\def\gge   {\gamma ^* \gamma  }

\def\Wgge {W_{\gamma \gamma}}
\def\Wgg {$W_{\gamma \gamma }$}
\def\Wvis {$W_{\mathrm{vis}}$}

\def\FQ2 {F$(Q^2)$}

\newcommand{\FF}{$F^\gamma_2$}
\newcommand{\FFE}{ F^\gamma_2 }

\def\qq    { \mathrm{q \bar{q}} }

\def\eeh { \ee \ra {} \ee ${\sl hadrons}$}
\def\eehe { \ee \ra {} \ee  {\sl hadrons} }
\def\Journal#1#2#3#4{{#1} {\bf #2} (#4) #3}

\def\NPB{{ Nucl. Phys. } B}

\def\PLB{{ Phys. Lett. } B}

\def\NIMA{{Nucl. Instr. Meth.} A}
\def\PRep{{ Phys. Rep. }}

\def\PRD{{Phys. Rev.} D}
\def\ZPC{{Z. Phys.} C}

\def\EURO{{ Eur. Phys. J. }}
\newlength{\capindent}
\newlength{\capwidth}
\newlength{\figwidth}
\newcommand{\icaption}[2][!*!,!]{\hspace*{\capindent}%
  \begin{minipage}{\capwidth}
    \ifthenelse{\equal{#1}{!*!,!}}%
      {\caption{#2}}%
      {\caption[#1]{#2}}
  \end{minipage}}
\newcommand{\be}{\begin{eqnarray}}
\newcommand{\eeq}{\end{eqnarray}}

\begin{document}
\begin{titlepage}
\title
{Measurement of the Photon Structure Function {\boldmath $ F^\gamma_2
 $ } \\ with the L3 Detector at LEP} \author{L3 Collaboration}

\begin{abstract}

The \eeh {} reaction, where one of the two electrons is detected in a
low polar-angle calorimeter, is analysed in order to measure the
hadronic photon structure function $F_{2}^{\gamma }$.  The full
high-energy and high-luminosity data set, collected with the L3
detector at centre-of-mass energies $189 \GeV \le \rts \le 209 \GeV $,
corresponding to an integrated luminosity of 608 pb$^{-1}$ is used.
The $ \Q2 $ range $ 11 \GeV^{2} \le \Q2 \le 34 \GeV^{2}$ and the $ x $
range $0.006 \le x \le 0.556$ are considered. The data are compared
with recent parton density functions.

\end{abstract}

\submitted

\end{titlepage}

\section{Introduction}
 
Photons are ideal tools for probing the structure of more complex
objects such as the proton in deep-inelastic scattering experiments.
At LEP, in the $\ee \ra {} \ee \gge \ra \ee {\sl hadrons}$ reaction,
two virtual photons are produced by the incoming
electrons\footnote{Throughout this Letter, the term ``electron"
indicates both electron and positron.} and their interaction yields
hadrons.  If the scattering angle of one of the electrons,
$\theta_{\mathrm{tag}}$, is sufficiently large, it is observed in the
low polar-angle electromagnetic BGO calorimeter~\cite{lumi} of the L3
detector~\cite{l3det}, originally devised to detect low angle Bhabha
scattering in order to measure the LEP luminosity.  This allows to
measure the four-momentum, $k'$, of this ``tagged'' electron. For
``single-tagged'' events the second electron is undetected, its polar
angle is small and the virtual photon radiated from this electron is
quasi-real.  In the framework of a deep-inelastic scattering formalism
the process \eeh {} is written as the convolution of the target photon
flux with the reaction $\mathrm{e} (k) + \gamma (p)\rightarrow
\mathrm{e}(k') + {\sl hadrons}$. The photon, $\gamma ^*$, with
four-momentum $q = k - k'$ and a large virtuality $ \Q2 = - \q2 \sim
{2E_{\mathrm{tag}} E_{\mathrm{beam}}(1-\cos\theta_{\mathrm{tag}})}$,
is considered as a point-like probe investigating the structure of the
target photon, $\gamma$, with four-momentum $p$ and virtuality $ \AP2
= - \ap2 \simeq 0 $.  $ E_{\mathrm{tag}}$ is the energy of the tagged
electron and $E_{\mathrm{beam}}$ the energy of the beam.  The
differential cross section is written in terms of the scaling
variables $x = \Q2 / 2(p \cdot q) = \Q2 / (\Q2 + \Wgge ^2 + \AP2 )$
and $y = (q \cdot p )/ (k \cdot p) = 1 - ( E_\mathrm{tag} /
E_\mathrm{beam} \cos ^2 \theta _\mathrm{tag} )$ as\cite{report ,
berger}:
\begin{equation}
\frac{{\mathrm d}\sigma _{{\mathrm e}\gamma \to {\mathrm eX}} (x,\Q2
)}{{\mathrm d}x{\mathrm d}\Q2 }=\frac{2\pi \alpha
^2}{xQ^4}[(1+(1-y)^2)F_2^\gamma (x,\Q2 )-y^2F_L^\gamma (x,\Q2 )]
\end{equation}
The variable $x$ depends on the two-photon centre-of-mass energy,
\Wgg, equal to the effective mass of the produced hadrons. The
inelasticity $y$ is small ($ y < 0.3$) in the kinematic region of this
study and consequently only $ \FFE (x, \Q2 )$ contributes appreciably
to the cross section. By convention, $\FFE / \alpha $ is measured,
where $\alpha $ is the fine-structure constant. Using this approach,
the photon structure function has been extensively studied at
low-energy \ee {} colliders~\cite{nisius} and at
LEP~\cite{alephopal,previous,nisius}.
    
A virtual photon can interact as a point-like particle in ``direct
processes"; it can fluctuate into a vector meson ($ \rho , \omega,
\phi $) inducing soft hadronic interactions in ``VDM processes" or it
can interact via its partonic content of quarks or gluons in
``resolved processes".  High $ \Q2 $ single-tag events favour
perturbative QED and QCD diagrams such as $\gamma \gamma \ra \qq $ , $
\gamma \mathrm{q} \ra \mathrm{g} \mathrm{q} $ or $\gamma \mathrm{g}
\ra \qq $. The two resolved processes, $ \gamma \mathrm{q} \ra
\mathrm{g} \mathrm{q} $ and $\gamma \mathrm{g} \ra \qq $, are
described using parton density functions extracted from the photon
structure functions measured in previous experiments at PEP, PETRA and
TRISTAN.  Reviews of the existing parameterisations may be found in
References~\citen{maria} and \citen{klasen}.  Recently, a new
parametrisation was obtained adding published LEP data\cite{cjk}.

This analysis uses the 608.1 \pb {} of high-energy LEP data, collected
at \ee {} centre-of-mass energies $189 \GeV \le \rts \le 209 \GeV $.
The data are grouped in four average \rts {} values, presented in
Table~\ref{table1}.  These high-energy data allow to extend our
previous measurements~\cite{previous} at $\rts \simeq 91 \GeV $ and
$\rts = 183 \GeV $ in the small-$x$ region down to 0.006 and in the
medium-$x$ region up to 0.556 for the $ \Q2 $ range $ 11 \GeV ^{2}\le
\Q2 \le 34 \GeV ^{2}$.  The \eeh {} cross section and the \FF {}
photon structure function are studied as a function of $x$ in the
three $ \Q2 $ intervals $11\GeV ^{2} \le \Q2 \le 14 \GeV^2$, $14\GeV
^{2} \le \Q2 \le 20 \GeV^2$ and $20\GeV ^{2} \le \Q2 \le 34 \GeV^2$.
The $ \Q2 $ evolution of \FF {} is also studied combining the values
at $ < \Q2 > = 12.4, 16.7$ and $25.5 \GeV ^{2}$ with our previous
measurements.

\section{Monte Carlo Models}
\label{sec:simulation}
The value of the $ \Q2 $ variable is accurately determined by
measuring the four-momentum of the scattered electron. However, the
effective mass of the final state hadrons is only partially
reconstructed, as these are often produced at low polar angles where
no tracking system can be installed. A Monte Carlo modelling of the
final state hadrons is therefore necessary\cite{lepwg} to determine
the $x$ variable.

Three Monte Carlo generators are used to model the process \eeh :
PHOJET\cite{PHOJET}, PYTHIA\cite{PYTHIA} and TWOGAM\cite{TWOGAM}.

  PHOJET describes hadron-hadron, photon-hadron and photon-photon
collisions. It is based on the Dual Parton Model combined with the
QCD-improved parton model~\cite{dpm}.  In order to have a continuous
transition between hard and soft processes, the distribution of the
transverse momentum, $ p_t$, of the soft partons is matched to the one
predicted by QCD.  The two-photon luminosity is calculated from the
flux of transversely polarised photons; corrections for the
longitudinally polarised photons are thus incorporated into an
effective two-photon cross section.  The transition from real-photon
to virtual-photon scattering is obtained by a change of the relative
weight of all partial cross sections.

PYTHIA is a general purpose Monte Carlo.  For two-photon interactions
it incorporates leading order (LO) hard-scattering processes as well
as elastic, diffractive and low $ p_t $ events. The classification of
the photon interactions into three different components, direct,
resolved and VDM, results in six different classes of events. Events
are also classified according to the hard scales involved in the
process: photon virtualities and parton transverse momenta.

TWOGAM generates three different processes separately: point-like
photon-photon interactions, resolved processes, and non-perturbative
soft processes described by the Generalised Vector Dominance Model
(GVDM).  The structure of the program is modular and the photon flux
is calculated with an exact LO formula.  The cross sections of the
three different processes are adjusted to fit the $x$-distribution of
the data. The cross section of the direct process is fixed to the
value expected in our kinematic range, $\sigma = 41$ pb.  The QCD
and the VDM cross sections are then adjusted to $\sigma = 5$ pb and
$\sigma = 28$ pb, respectively.

For the three Monte Carlo generators parton showering and
hadronisation are described by JETSET\cite{jetset}.  The dominant
backgrounds are evaluated with PYTHIA for $\ee \to \qq (\gamma)$ and
DIAG36 \cite{diag36} for $ \ee \to \ee \tau ^+ \tau ^- $.
 
All Monte Carlo samples are generated with a luminosity at least five
times greater than the experimental one. All events are passed through
a full detector simulation which uses the GEANT\cite{geant} and
GHEISHA\cite{geisha} programs and takes into account detector
efficiencies and time-dependent effects. Monte Carlo events are then
reconstructed in the same way as the data.

\section{Data Analysis}

Events are mainly accepted by two independent triggers: the single-tag
trigger and the central track-trigger.  The single-tag trigger
requires at least 70\% of the beam energy to be deposited in one of
the low polar-angle calorimeters, in coincidence with at least one
track in the central tracking chamber.  The central track-trigger
requires at least two tracks back-to-back in the transverse plane
within $\pm 60^\circ $, each with $p_{t} >$ 150 \MeV .  The average
trigger efficiency is about 97\%. There is a single-tag trigger signal
for about 90\% of the selected events and a central track-trigger
signal for about 85\% of the selected events.

\subsection{Event Selection }

\begin{figure}[h!]
\begin{center}
\includegraphics[width=0.45\textwidth]{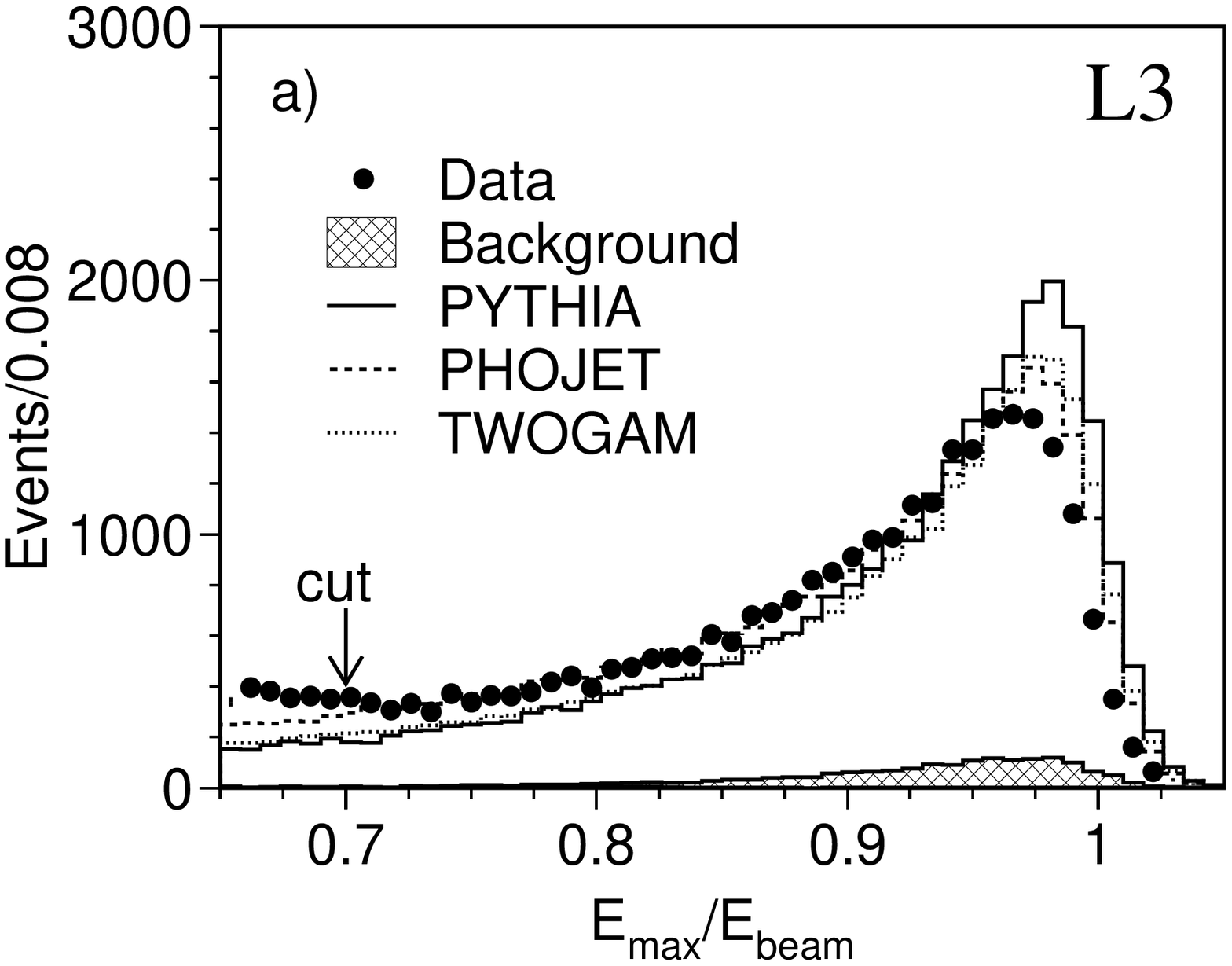}
\includegraphics[width=0.45\textwidth]{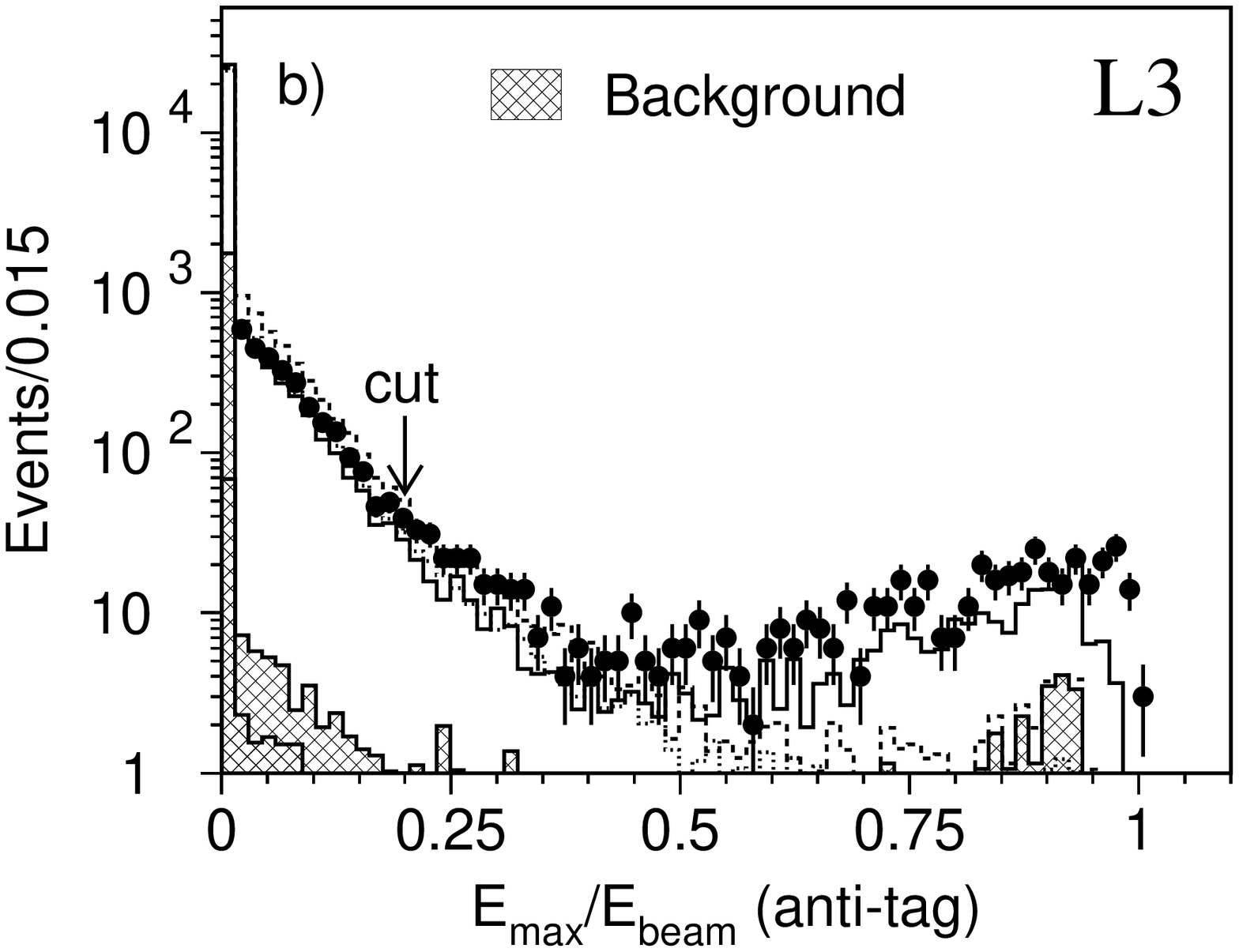}
\includegraphics[width=0.45\textwidth]{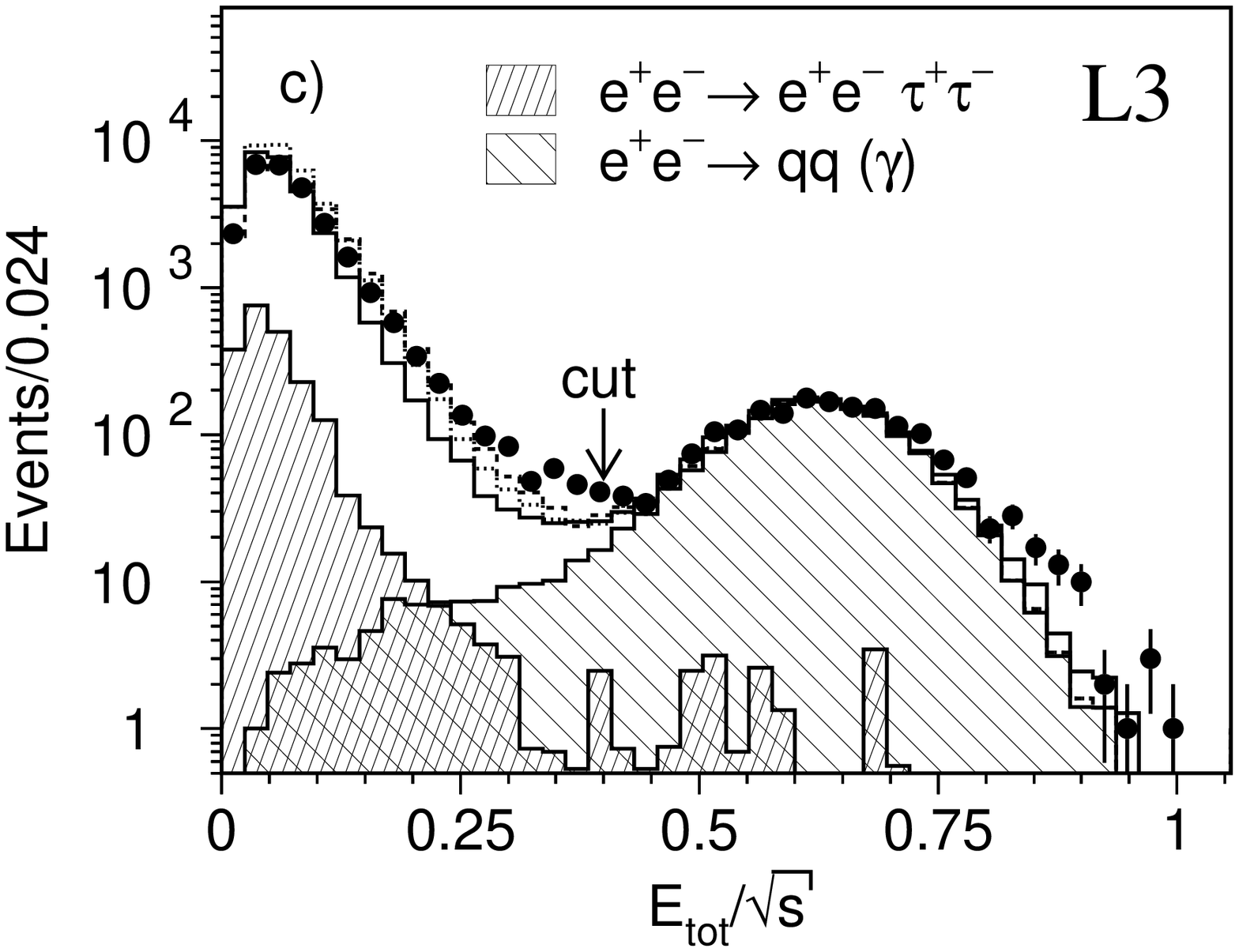} 
\includegraphics[width=0.45\textwidth]{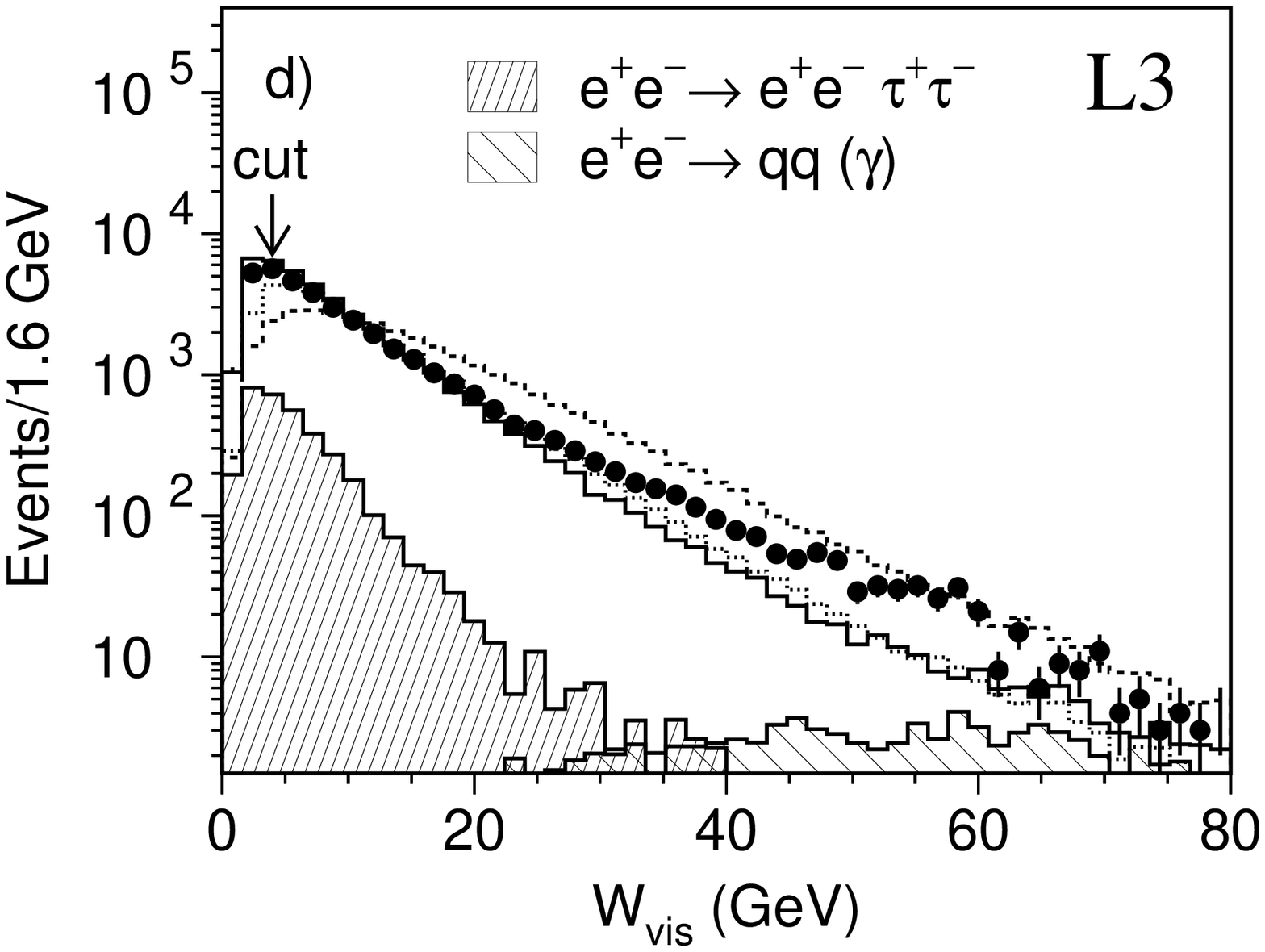}
\end{center}
\caption{Distribution of the highest energy clusters in the forward
electromagnetic calorimeters for a) the tagged electron side and b)
for the opposite side.  c) Total energy in the central
calorimeters. d) The visible mass of the hadronic final state.  All
distributions are presented after all other cuts are applied.  The
backgrounds from annihilation and two-photon leptonic events are
indicated as shaded areas and added to the expectations of the PYTHIA,
PHOJET and TWOGAM generators.  The arrows indicate the position of the
cuts.}
 
\label{fig1}
\end{figure}

Events are selected by requiring a single scattered electron in the
low polar-angle calorimeter and a hadronic final state.  A tagged
electron candidate is the highest energy cluster with a shape
consistent with an electromagnetic shower, $E_{\mathrm{tag}} /
E_{\mathrm{beam}} > 0.7$, as shown in Figure~\ref{fig1}a, and a polar
angle in the fiducial region $0.0325 \;\mathrm{rad} \le \theta \le
0.0637 \;\mathrm{rad}$ inside the geometrical acceptance $0.030
\;\mathrm{rad} \le \theta \le 0.066 \;\mathrm{rad}$.  To ensure that
the virtuality of the target photon is small, the highest-energy
cluster in the low polar-angle calorimeter opposite to the tagged
electron must have an energy less than $20{\%}$ of the beam energy, as
shown in Figure~\ref{fig1}b.
 
At least four additional particles must be detected.  A particle can
be a track or a photon. A track must have $p_{t} > 100 \MeV $ and a
distance of closest approach in the transverse plane to the
interaction vertex of less than 10 mm. A photon is a cluster in the
electromagnetic BGO calorimeters with energy above 100 \MeV , not
associated with a charged track.

To reduce the background from the process $\ee \to \qq (\gamma )$, the
total energy deposited in the electromagnetic and hadronic
calorimeters must be less than 40{\%} of the center-of-mass energy, as
shown in Figure~\ref{fig1}c.  The events with a large value of the
total energy are due to the $\ee\to \mathrm{Z} \gamma \to \qq\gamma $
process, where the radiative photon is misidentified as the tagged
electron.

The mass of the hadronic final state, \Wvis, is calculated from all
tracks and calorimetric clusters. Additional clusters detected in the
low polar-angle calorimeter are assigned the pion mass and included in
the calculation of \Wvis.  To avoid the hadronic resonance region,
\Wvis\ is required to be greater than 4 \GeV, as presented in
Figure~\ref{fig1}d.
 
\begin{figure}[h!]
\begin{center}
\includegraphics[width=0.45\textwidth]{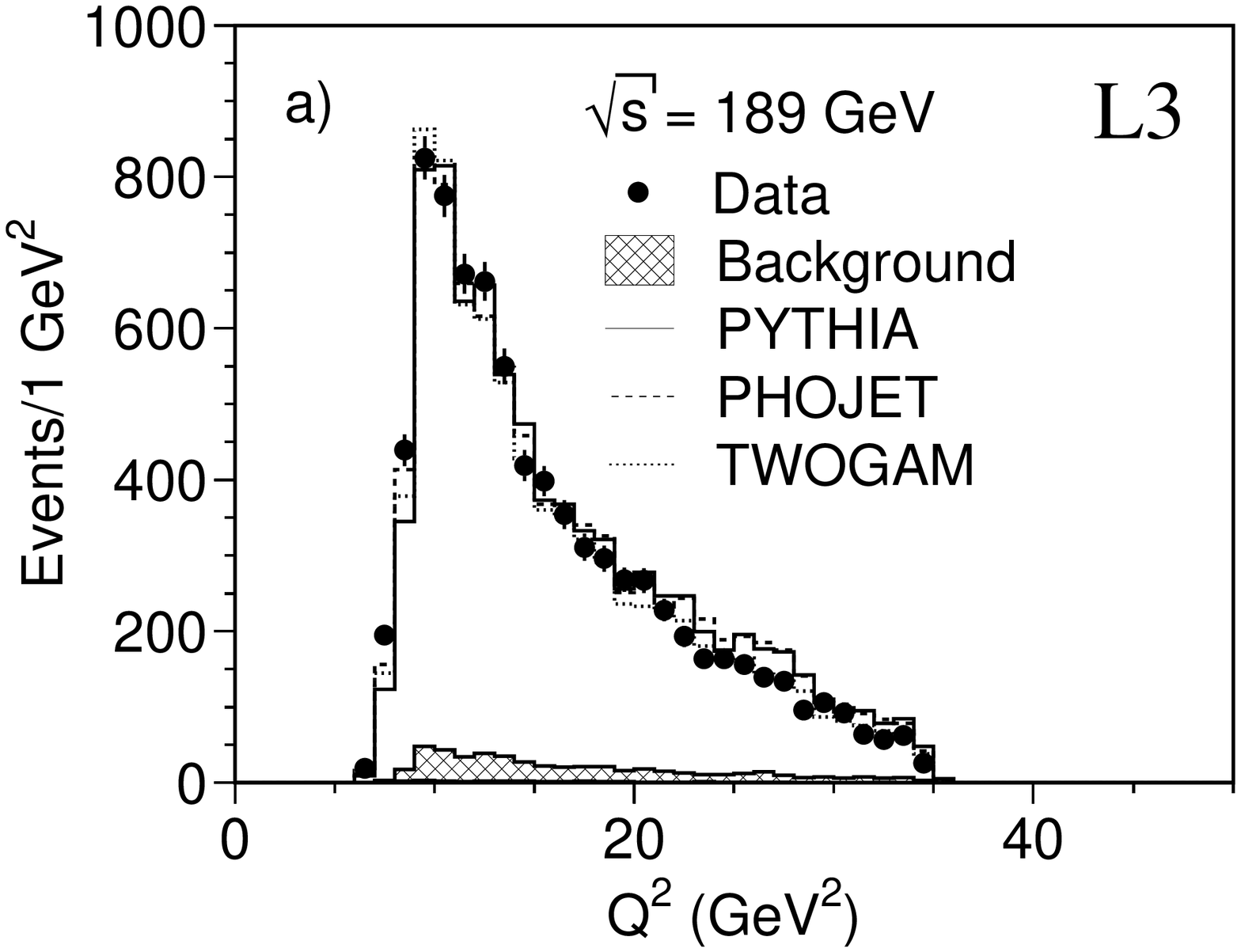}
\includegraphics[width=0.45\textwidth]{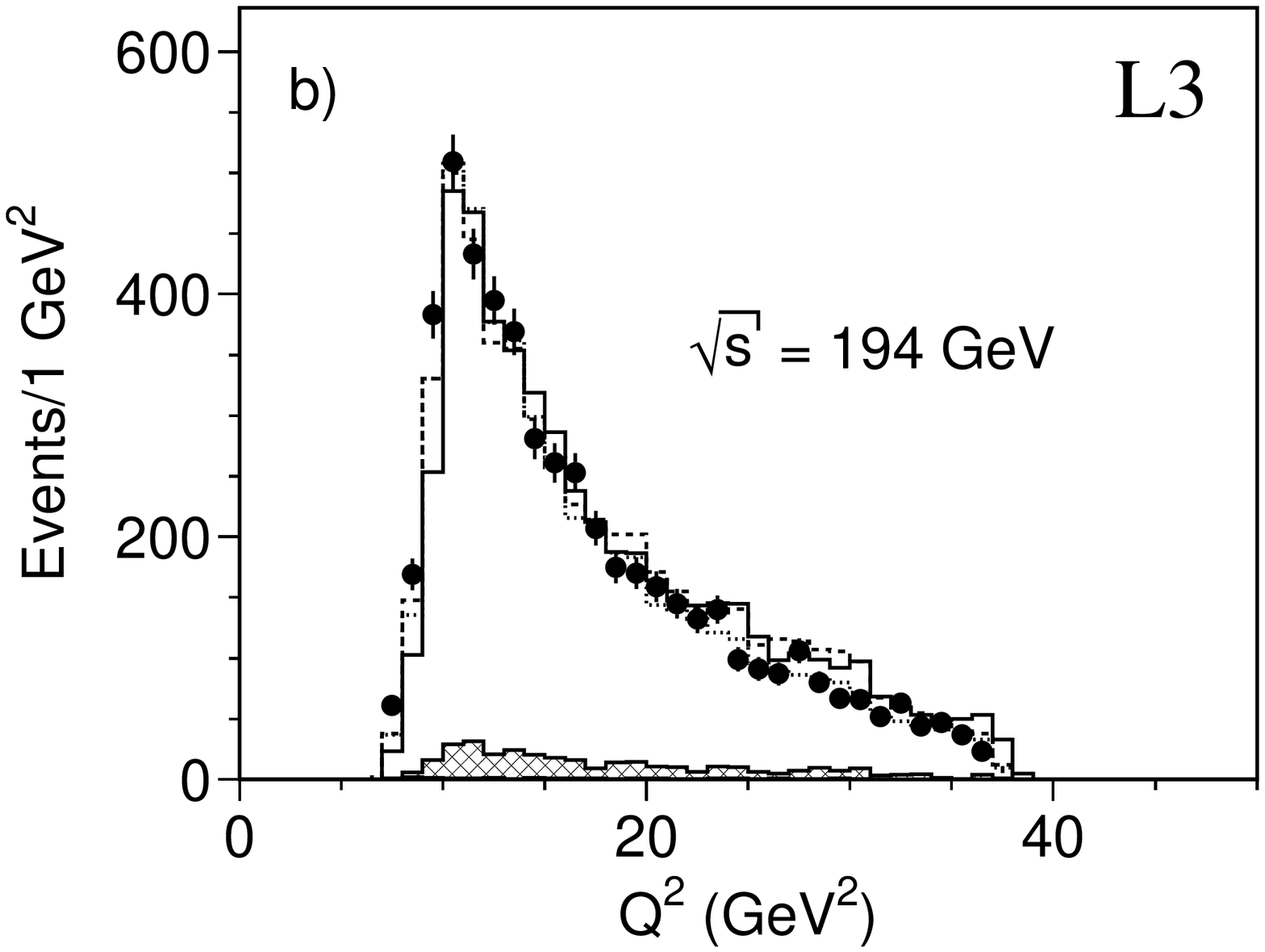}
\includegraphics[width=0.45\textwidth]{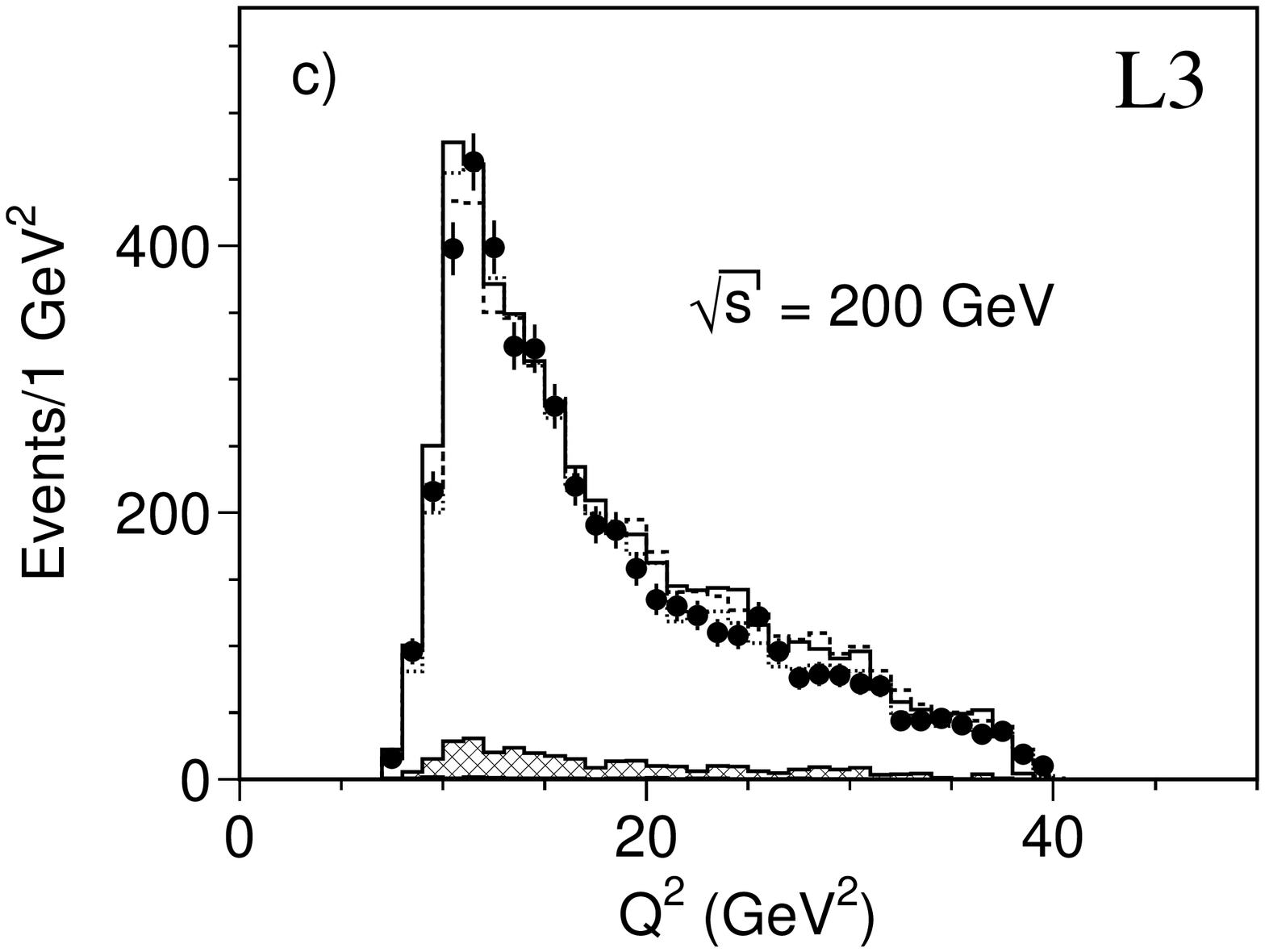}
\includegraphics[width=0.45\textwidth]{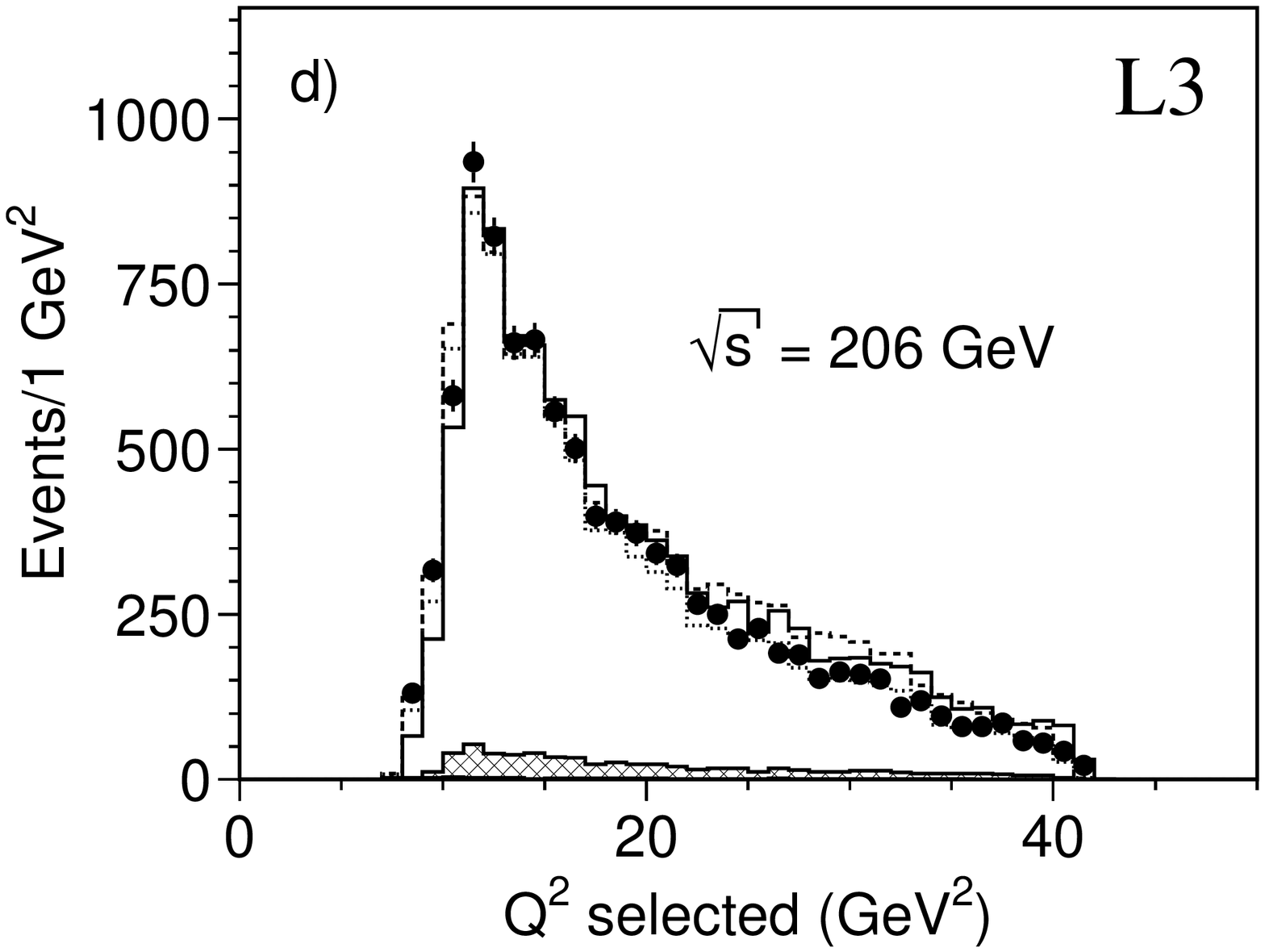}
\end{center}
\caption{$ \Q2 $ distribution of the selected events for the four
average \rts {} ranges.  }
\label{fig2}
\end{figure}

\begin{figure}[htbp]
\begin{center}
\includegraphics[width=0.7\textwidth]{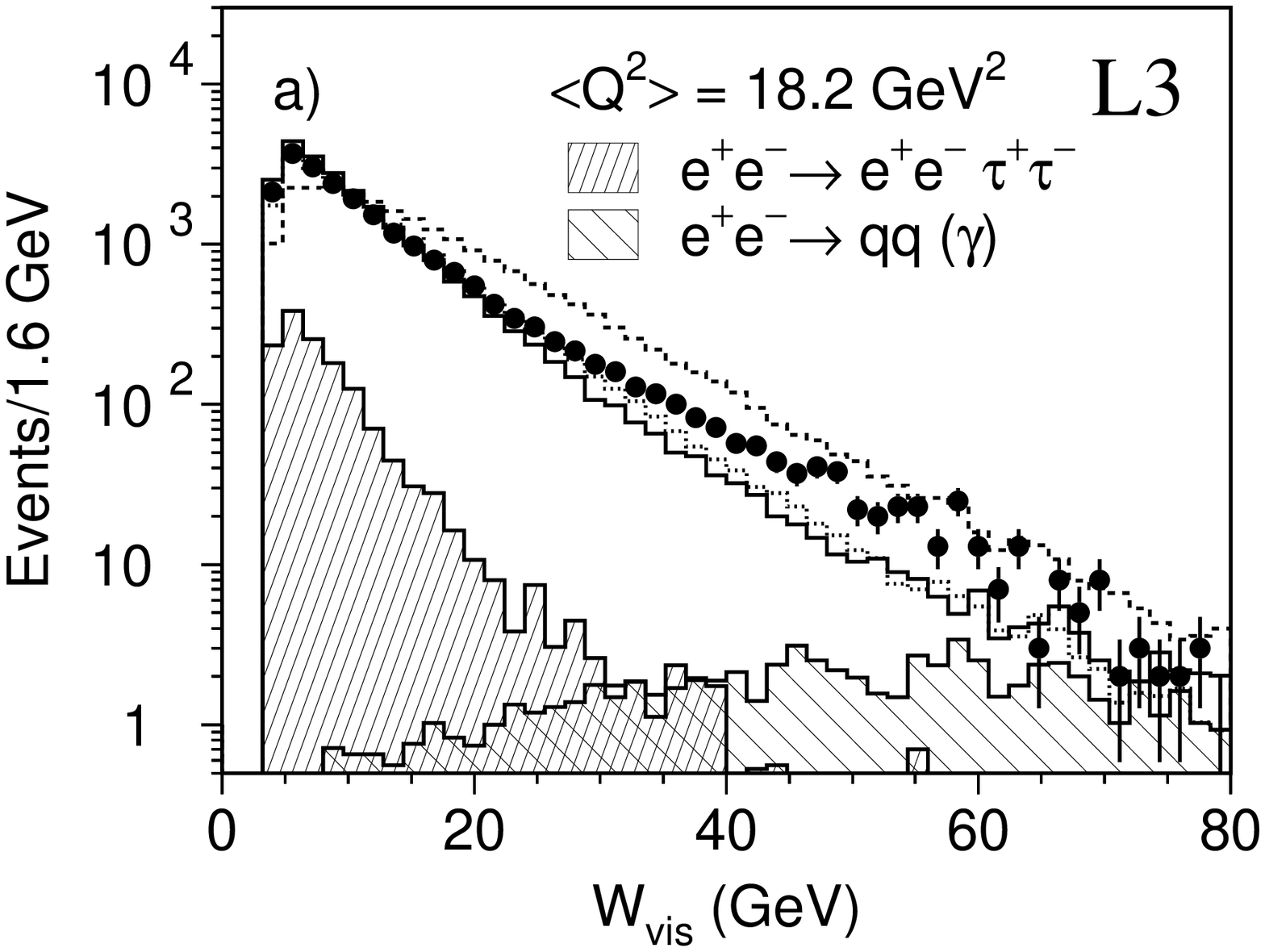}
\includegraphics[width=0.7\textwidth]{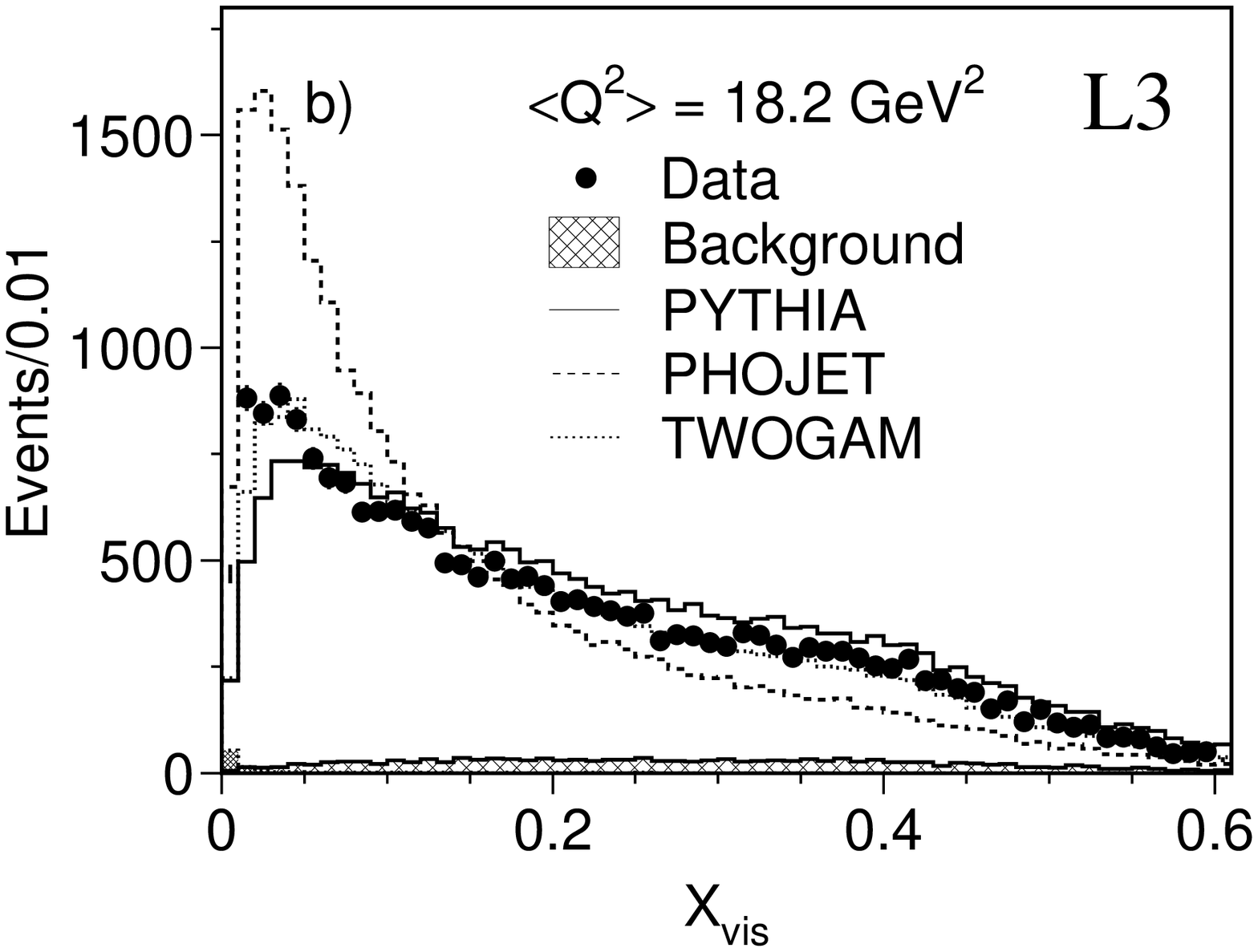}
\end{center}
\caption{Distribution of the visible mass of the two-photon system and
of the $x_{\rm vis}$ for all selected events compared with Monte Carlo
predictions. The backgrounds from annihilation and two-photon leptonic
events are indicated as shaded areas and added to the expectations of
the PYTHIA, PHOJET and TWOGAM generators.  }
\label{fig3}
\end{figure}

Figure~\ref{fig2} shows the $\Q2 $ distribution for each \rts {}
sample.  Only events with $11 \GeV ^2 \le \Q2 \le 34 \GeV ^2$ are
studied.  The number of selected events and the backgrounds from the $
\ee \to \ee \tau ^+ \tau ^- $ and $\ee \to \qq (\gamma )$ processes
are given in Table~\ref{table1}.  The background is dominated by the $
\ee \to \ee \tau ^+ \tau ^- $ production.  The contribution from the
$\ee \to \tau^+ \tau^-$ and $\ee \to \W^+ \W^-$ processes is
negligible. The background from beam-gas and beam-wall events is found
to be negligible by inspection of the radial distribution of track
intersections.  The \Wvis {} and $x_{\mathrm{vis}} ={ \Q2 }
\mathord{\left/ {\vphantom {{ \Q2 } {( \Q2 + {W}_{\mathrm{vis}}^2 )}}}
\right.  \kern-\nulldelimiterspace} ({ \Q2 + W_{\mathrm{vis}}^2) }$
distributions are presented in Figure~\ref{fig3} for all selected
data.  The PYTHIA and TWOGAM model reproduce the data rather well,
except at large values of ${W}_{\mathrm{vis}}$.  PHOJET presents a
harder mass spectrum and predicts too many events for
$x_{\mathrm{vis}} < 0.1$ and is therefore not used in the following.

The total acceptance is calculated for each data sample separately. It
takes into account the trigger efficiency, the geometrical acceptance
and the selection cuts.  An example is presented in Figure 4a for the
data at \rts = 189 \GeV .
 
\section{Results}
\subsection{Unfolding and Differential Cross Sections }

The cross section $\Delta \sigma / \Delta x $ as a function of $x$ for
the reaction \eeh {} is measured for three $\Q2 $ intervals: $11
\GeV^2 \le \Q2 \le 14 \GeV^2$, $14 \GeV^2 \le \Q2 \le 20 \GeV^2$ and
$20 \GeV^2 \le \Q2 \le 34 \GeV^2$.  Each data set is subdivided into
bins of $x_\mathrm{vis}$ of similar statistics, as listed in
Table~\ref{table2}.  A Bayesian unfolding procedure\cite{bayes} is
used to relate the measured $x_\mathrm{vis}$ to the true value of $x$
and to correct the data for the detector acceptance and efficiency.
This procedure is applied using, in turn, the PYTHIA and TWOGAM
generators. The average of the cross sections obtained in the two
cases is retained.  The correlation between the generated value of
$x$ and $x_\mathrm{vis}$ is similar for the two models. The one
obtained with PYTHIA is shown in Figure 4b.  The cross sections
measured for each $x$ interval of average value $\langle x\rangle$ and
for each value of \rts {} are given in Table~\ref{table2} with their
statistical and systematic uncertainties. The bin-to-bin correlation
matrices obtained with PYTHIA and TWOGAM for $\rts=189\GeV$ are shown
in Tables~\ref{table2a} and~\ref{table2abis}, respectively. Similar
matrices are obtained for the other values of \rts.

\subsection{Systematic Uncertainties} 

The systematic uncertainties on the cross sections are estimated for
each data sample, for each $x$ bin and for each $\Q2 $ interval. Three
main sources of systematic uncertainties are considered: the selection
procedure, the trigger efficiency and the Monte Carlo model. Their
average effects over the full data sample are listed in Table~\ref{table3}.

The uncertainties from the selection procedure are estimated by
varying the selection cuts. The fiducial value of the polar angle in
the low polar-angle calorimeter is varied from $0.0325 \;
\mathrm{rad}$ to $0.0360 \;\mathrm{rad}$ and from $ 0.0637
\;\mathrm{rad}$ to $ 0.060\; \mathrm{rad}$.  These changes result in a
$\Q2 $-dependent uncertainty, as the highest and lowest values of the
$\Q2 $ are affected by the fiducial cut.  The cut on $E_{\mathrm{tag}}
$ is varied from $0.70E_{\mathrm{beam}}$ to $0.65E_{\mathrm{beam}}$
and $0.75E_{\mathrm{beam}}$.  The anti-tag cut is changed from $0.20
E_{\mathrm{beam}}$ to $0.15 E_{\mathrm{beam}}$ and $0.25
E_{\mathrm{beam}}$. The numbers of particles is varied from four to
three and five.  The cut on the total energy in the calorimeters
relative to \rts {} is varied from 0.40 to 0.35 and 0.45.  The
uncertainty on the trigger efficiency, as determined from the data, is
1.5\%.  An additional uncertainty comes from the limited Monte Carlo statistics.

The discrepancies of the results obtained with the PYTHIA and TWOGAM
generators are considered as systematic uncertainties related to the
Monte Carlo modelling.  This difference is due to the calculated
acceptance as well as to the unfolding procedure. An additional
contribution to this modelling uncertainty is evaluated by repeating
the analysis doubling or halving the VDM component of the TWOGAM
generator and is found to be negligible.

\subsection{Extraction of {\boldmath $  F^\gamma_2 $ }   }

The photon structure function \FF /$\alpha $ is derived from the cross
section of Equation 1 and the target-photon flux calculated by the
program GALUGA\cite{galuga} as:
\[
F_2^\gamma (x,Q^2)/ \alpha = \frac{\Delta\sigma _\mathrm{meas} ( \eehe
)}{\Delta\sigma _\mathrm{Galuga} ( \eehe ) }
\]
The program calculates the theoretical value $\Delta\sigma
_\mathrm{Galuga}$ in the given $ \Q2 $ and $x$ range setting
$F_2^\gamma =1$ and $F_L^\gamma $ to the QPM value\cite{walsh}.  In
practice the $F_L^\gamma $ contribution to the cross section is
smaller than 1\%, due to the small value of $y$. The running of the
fine-structure constant with $\Q2 $ is included.  A GVDM form factor
is used in the calculation for the target photon virtuality whose
average value is of the order of $ 0.07 \GeV ^2$.  The low polar-angle
calorimeter acceptance for the tagged and the anti-tagged electron and
the $ \Wgge > 4 \GeV $ requirement are taken into account.  The
uncertainty on $\Delta\sigma _\mathrm{Galuga}$, as estimated by
comparing the GVDM to a $\rho$ form factor, is 2\%.
 
The contribution of radiative corrections to the cross section is
evaluated by using the program RADCOR\cite{radcor}, which includes
initial and final state radiation for the reaction $ \ee \ra \ee \mm
$. The corrections are mainly due to initial state radiation from the
electron scattered at large angle. Final state radiation is detected
together with the scattered electron due to the finite granularity of
the calorimeter. Initial state radiation from the electron producing
the quasi-real target photon is very small. The calculations are
performed at the generator level using the $ \Q2 $ from the electron
variables and \Wgg {} from the muon pair. The measured \FF /$\alpha $
is multiplied by the ratio, $\cal R$, of the non-radiative and the
total cross section, shown in Tables~\ref{table4} and \ref{table5} for
different values of $x$ and $\Q2 $.

The \FF {} values, averaged over the $x$ intervals, are first obtained
for each individual data set. The results are statistically compatible
and, consequently, a weighted average of \FF {} is calculated for the
$\Q2 $ ranges with average values $ 12.4\GeV ^{2}, 16.7\GeV ^{2}$ and
$25.5 \GeV ^{2}$.  This procedure is applied to data unfolded
separately with PYTHIA and TWOGAM and the two different values are
shown in Figure~\ref{fig5}.  Their average value is given in
Table~\ref{table4} and in Figure~\ref{fig6} for each $x$ interval of
expected average value $\langle x \rangle$.  In addition to the
systematic uncertainty on the cross section, presented in
Table~\ref{table3}, two systematic uncertainties are further
considered in the extraction of \FF. The first uncertainty, of $2\% $,
is associated to the GALUGA calculations. The second
uncertainty, also of $2 \% $, covers the estimation of the initial-state
and final-state radiative corrections. The uncertainty on
initial-state radiation is assessed by changing the angular and
momentum criteria which separate soft from hard photons in the Monte
Carlo programs. The uncertainty on final-state radiation is estimated
by varying the cone angle of the calorimeter for which final state
radiation is detected together with the scattered electron.
 
A comparison of the data with the existing parameterisations as
obtained with the PDFLIB library~\cite{pdflib} shows that our data are
not well described by the leading-order parton density functions.  In
Figure~\ref{fig6} the data are compared with the predictions of the
leading- and higher-order parton density functions GRV~\cite{GRV} and
the higher-order parton density functions CJK~\cite{cjk}. The best
agreement is found for the higher-order GRV~\cite{GRV} predictions.
In all cases four quarks, u, d, s and c are used.  The pure QPM
prediction for $\gamma \gamma \to \qq $ is also indicated, as
calculated by using GALUGA with a mass of 0.32 \GeV {} for the u and d
quarks, 0.5 \GeV {} for the s quark and 1.4 \GeV {} for the c
quark. It is clearly insufficient to describe the data.
 
\subsection{{\boldmath $ \Q2 $}-evolution of {\boldmath $F_{2}^{\gamma }$ }  }

The $ \Q2 $-evolution of \FF , is studied in four $x$ bins,
$x=0.01-0.1$, $x=0.1-0.2$, $x=0.2-0.3$ and $x=0.3-0.5$ and the results
are given in Table~\ref{table5}.  In Figure~\ref{fig7} the $
F_2^\gamma/\alpha $ values are presented for the lowest $x$ bin and
for a combined bin $x= 0.1 - 0.5$, together with our previous
results~\cite{previous}.  Corrections for radiative effects are
applied.  The new measurements at $11 \GeV ^2\le \Q2 \le 14 \GeV ^2$,
$14 \GeV ^2\le \Q2 \le 20 \GeV ^2$ and $20\GeV ^2 \le \Q2 \le 34 \GeV
^2$ are in good agreement with our previous results.  The expected
linear growth with $ \ln \Q2 $ is observed in both $x$ intervals. The
function $a + b \ln ({Q^2}/\GeV^2)$ is fitted to the data, taking into
account the total uncertainty calculated from the quadratic sum of
statistical and systematic uncertainties. The fit results are: $a =
0.141 \pm 0.007$ and $b = 0.060 \pm 0.005$ for $ x=0.01- 0.1 $ with a
confidence level of 44\% and $a = 0.05\pm 0.11 $ and $ b = 0.13 \pm
0.04$ for $ x = 0.1- 0.5 $ with a confidence level of 71\%.
   
The predictions of the leading- and higher-order parton density
functions GRV~\cite{GRV} and the higher-order parton density functions
CJK~\cite{cjk} are also indicated in Figure~\ref{fig7}.  The evolution
is different for the different models and the data are better
described by the higher-order GRV model.

\section{Conclusions}

The photon structure function \FF {} is measured at LEP with the L3
detector at centre-of-mass energies $189 \le \rts \le 209 \GeV $ in
the $ \Q2 $ range $ 11 \GeV ^2 \le \Q2 \le 34 \GeV ^2 $ and the $x$
range $ 0.006 \le x \le 0.556$.  The data are better reproduced by the
higher-order parton density function of GRV than by other parton
distribution functions determined from the low energy data.

Combining the present results with previous L3 measurements, the $\Q2
$ evolution is studied from $1.5 \GeV ^2$ to $120 \GeV ^2$ in the
low-$x$ region, $ 0.01 \le x \le 0.1 $, and from $12.4 \GeV ^2$ to $
225 \GeV ^2$ in the higher-$x$ region, $ 0.1 < x \le 0.5 $. The
measurements at different centre-of-mass energies are consistent and
the $\ln \Q2 $ evolution of \FF {} is clearly confirmed.

%
%
%
\newpage
\typeout{   }     
\typeout{Using author list for paper 287 -  }
\typeout{$Modified: Jul 15 2001 by smele $}
\typeout{!!!!  This should only be used with document option a4p!!!!}
\typeout{   }
%
%
%
%
%
%

\newcount\tutecount  \tutecount=0
\def\tutenum#1{\global\advance\tutecount by 1 \xdef#1{\the\tutecount}}
\def\tute#1{$^{#1}$}
\tutenum\aachen            
\tutenum\nikhef            
\tutenum\mich              
\tutenum\lapp              
\tutenum\basel             
\tutenum\lsu               
\tutenum\beijing           
\tutenum\bologna           
\tutenum\tata              
\tutenum\ne                
\tutenum\bucharest         
\tutenum\budapest          
\tutenum\mit               
\tutenum\panjab            
\tutenum\debrecen          
\tutenum\dublin            
\tutenum\florence          
\tutenum\cern              
\tutenum\wl                
\tutenum\geneva            
\tutenum\hamburg           
\tutenum\hefei             
\tutenum\lausanne          
\tutenum\lyon              
\tutenum\madrid            
\tutenum\florida           
\tutenum\milan             
\tutenum\moscow            
\tutenum\naples            
\tutenum\cyprus            
\tutenum\nymegen           
\tutenum\caltech           
\tutenum\perugia           
\tutenum\peters            
\tutenum\cmu               
\tutenum\potenza           
\tutenum\prince            
\tutenum\riverside         
\tutenum\rome              
\tutenum\salerno           
\tutenum\ucsd              
\tutenum\sofia             
\tutenum\korea             
\tutenum\taiwan            
\tutenum\tsinghua          
\tutenum\purdue            
\tutenum\psinst            
\tutenum\zeuthen           
\tutenum\eth               

{
\parskip=0pt
\noindent
{\bf The L3 Collaboration:}
\ifx\selectfont\undefined
 \baselineskip=10.8pt
 \baselineskip\baselinestretch\baselineskip
 \normalbaselineskip\baselineskip
 \ixpt
\else
 \fontsize{9}{10.8pt}\selectfont
\fi
\medskip
\tolerance=10000
\hbadness=5000
\raggedright
\hsize=162truemm\hoffset=0mm
\def\r{\rlap,}
\noindent

P.Achard\r\tute\geneva\ 
O.Adriani\r\tute{\florence}\ 
M.Aguilar-Benitez\r\tute\madrid\ 
J.Alcaraz\r\tute{\madrid}\ 
G.Alemanni\r\tute\lausanne\
J.Allaby\r\tute\cern\
A.Aloisio\r\tute\naples\ 
M.G.Alviggi\r\tute\naples\
H.Anderhub\r\tute\eth\ 
V.P.Andreev\r\tute{\lsu,\peters}\
% ...
\newpage
\begin{list}{A}{\itemsep=0pt plus 0pt minus 0pt\parsep=0pt plus 0pt minus 0pt
                \topsep=0pt plus 0pt minus 0pt}
\item[\aachen]
 III. Physikalisches Institut, RWTH, D-52056 Aachen, Germany$^{\S}$
\item[\nikhef] National Institute for High Energy Physics, NIKHEF, 
     and University of Amsterdam, NL-1009 DB Amsterdam, The Netherlands
\item[\mich] University of Michigan, Ann Arbor, MI 48109, USA
\item[\lapp] Laboratoire d'Annecy-le-Vieux de Physique des Particules, 
     LAPP,IN2P3-CNRS, BP 110, F-74941 Annecy-le-Vieux CEDEX, France
\item[\basel] Institute of Physics, University of Basel, CH-4056 Basel,
     Switzerland
\item[\lsu] Louisiana State University, Baton Rouge, LA 70803, USA
\item[\beijing] Institute of High Energy Physics, IHEP, 
  100039 Beijing, China$^{\triangle}$ 
\item[\bologna] University of Bologna and INFN-Sezione di Bologna, 
     I-40126 Bologna, Italy
\item[\tata] Tata Institute of Fundamental Research, Mumbai (Bombay) 400 005, India
\item[\ne] Northeastern University, Boston, MA 02115, USA
\item[\bucharest] Institute of Atomic Physics and University of Bucharest,
     R-76900 Bucharest, Romania
\item[\budapest] Central Research Institute for Physics of the 
     Hungarian Academy of Sciences, H-1525 Budapest 114, Hungary$^{\ddag}$
\item[\mit] Massachusetts Institute of Technology, Cambridge, MA 02139, USA
\item[\panjab] Panjab University, Chandigarh 160 014, India
\item[\debrecen] KLTE-ATOMKI, H-4010 Debrecen, Hungary$^\P$
\item[\dublin] Department of Experimental Physics,
  University College Dublin, Belfield, Dublin 4, Ireland
\item[\florence] INFN Sezione di Firenze and University of Florence, 
     I-50125 Florence, Italy
\item[\cern] European Laboratory for Particle Physics, CERN, 
     CH-1211 Geneva 23, Switzerland
\item[\wl] World Laboratory, FBLJA  Project, CH-1211 Geneva 23, Switzerland
\item[\geneva] University of Geneva, CH-1211 Geneva 4, Switzerland
\item[\hamburg] University of Hamburg, D-22761 Hamburg, Germany
\item[\hefei] Chinese University of Science and Technology, USTC,
      Hefei, Anhui 230 029, China$^{\triangle}$
\item[\lausanne] University of Lausanne, CH-1015 Lausanne, Switzerland
\item[\lyon] Institut de Physique Nucl\'eaire de Lyon, 
     IN2P3-CNRS,Universit\'e Claude Bernard, 
     F-69622 Villeurbanne, France
\item[\madrid] Centro de Investigaciones Energ{\'e}ticas, 
     Medioambientales y Tecnol\'ogicas, CIEMAT, E-28040 Madrid,
     Spain${\flat}$ 
\item[\florida] Florida Institute of Technology, Melbourne, FL 32901, USA
\item[\milan] INFN-Sezione di Milano, I-20133 Milan, Italy
\item[\moscow] Institute of Theoretical and Experimental Physics, ITEP, 
     Moscow, Russia
\item[\naples] INFN-Sezione di Napoli and University of Naples, 
     I-80125 Naples, Italy
\item[\cyprus] Department of Physics, University of Cyprus,
     Nicosia, Cyprus
\item[\nymegen] Radboud University and NIKHEF, 
     NL-6525 ED Nijmegen, The Netherlands
\item[\caltech] California Institute of Technology, Pasadena, CA 91125, USA
\item[\perugia] INFN-Sezione di Perugia and Universit\`a Degli 
     Studi di Perugia, I-06100 Perugia, Italy   
\item[\peters] Nuclear Physics Institute, St. Petersburg, Russia
\item[\cmu] Carnegie Mellon University, Pittsburgh, PA 15213, USA
\item[\potenza] INFN-Sezione di Napoli and University of Potenza, 
     I-85100 Potenza, Italy
\item[\prince] Princeton University, Princeton, NJ 08544, USA
\item[\riverside] University of Californa, Riverside, CA 92521, USA
\item[\rome] INFN-Sezione di Roma and University of Rome, ``La Sapienza",
     I-00185 Rome, Italy
\item[\salerno] University and INFN, Salerno, I-84100 Salerno, Italy
\item[\ucsd] University of California, San Diego, CA 92093, USA
\item[\sofia] Bulgarian Academy of Sciences, Central Lab.~of 
     Mechatronics and Instrumentation, BU-1113 Sofia, Bulgaria
\item[\korea]  The Center for High Energy Physics, 
     Kyungpook National University, 702-701 Taegu, Republic of Korea
\item[\taiwan] National Central University, Chung-Li, Taiwan, China
\item[\tsinghua] Department of Physics, National Tsing Hua University,
      Taiwan, China
\item[\purdue] Purdue University, West Lafayette, IN 47907, USA
\item[\psinst] Paul Scherrer Institut, PSI, CH-5232 Villigen, Switzerland
\item[\zeuthen] DESY, D-15738 Zeuthen, Germany
\item[\eth] Eidgen\"ossische Technische Hochschule, ETH Z\"urich,
     CH-8093 Z\"urich, Switzerland
\item[\S]  Supported by the German Bundesministerium 
        f\"ur Bildung, Wissenschaft, Forschung und Technologie.
\item[\ddag] Supported by the Hungarian OTKA fund under contract
numbers T019181, F023259 and T037350.
\item[\P] Also supported by the Hungarian OTKA fund under contract
  number T026178.
\item[$\flat$] Supported also by the Comisi\'on Interministerial de Ciencia y 
        Tecnolog{\'\i}a.
\item[$\sharp$] Also supported by CONICET and Universidad Nacional de La Plata,
        CC 67, 1900 La Plata, Argentina.
\item[$\triangle$] Supported by the National Natural Science
  Foundation of China.
\end{list}
}
\vfill


\newpage

\begin{table}[htb]
\begin{center}
\begin{tabular}{|c|c|c|c|c|c|}
\hline
\rule{0pt}{12pt}
$ \langle \rts \rangle $(\GeV) & $\cal{L} ( \pb )$ & Events & $\ee \to \ee \tau^+ \tau^-$ & $\ee \to \qq (\gamma )$ & Purity (\%)\\

\hline
 \multicolumn{6} {|c|}{$Q^2=11-14 \GeV ^2$ }\\
\hline  

189 & 171.8 & 1884 & 107.6 & \phantom{0}5.7& 94\\
194 & 111.4 & 1197 & \phantom{0}76.3  & \phantom{0}3.3& 93\\
200 & 109.3 & 1187 & \phantom{0}74.9  & \phantom{0}3.7& 93\\
206 & 215.6 & 2418 & 129.6 & \phantom{0}7.7& 94\\

\hline
 \multicolumn{6} {|c|}{$Q^2=14-20 \GeV ^2$}\\
\hline  

189 & 171.8 & 2046 & 128.6 & \phantom{0}9.7 & 93\\
194 & 111.4 & 1347 & \phantom{0}91.2  & \phantom{0}5.4 & 93\\
200 & 109.3 & 1359 & \phantom{0}89.5  & \phantom{0}4.9 & 93\\
206 & 215.6 & 2886 & 177.4 & \phantom{0}8.7 & 94\\

\hline
 \multicolumn{6} {|c|}{$Q^2=20-34 \GeV ^2$ }\\
\hline  

189 & 171.8 & 1922 & 143.9 & \phantom{0}8.1 & 92\\
194 & 111.4 & 1331 & 103.6 & \phantom{0}5.8 & 92\\
200 & 109.3 & 1287 & 101.6 & \phantom{0}6.7 & 92\\
206 & 215.6 & 2859 & 202.5 & 12.7& 92\\

\hline
\end{tabular}
\end{center}

\caption{ The average  \ee {} centre-of-mass energies, $  \langle \rts \rangle $, and the
corresponding  luminosities 
for the four data samples  together with  the number of selected events  in the $ \Q2 $ 
intervals  $11 \GeV^2 \le \Q2 \le 14 \GeV^2$, $14\GeV^2 \le \Q2 \le 20 \GeV^2$
and $20 \GeV^2\le \Q2 \le 34 \GeV^2$. The numbers of expected  background events
from the $\ee \to \ee \tau^+ \tau^-$ and $\ee \to \qq (\gamma )$ processes and the signal purity are also listed.}

\label{table1}
\end{table}

 \hspace*{-0.5cm}
\begin{table}{\small \begin{center} 
\begin{tabular}{|c|c|c|c|c|c|}
\hline
 \multicolumn{2} {|c|} {$\langle \sqrt{s} \rangle $ } &  $189   \GeV $ & $194  \GeV $& $200  \GeV $&$ 206  \GeV $\\
\hline
\multicolumn{6} {|c|}{$Q^2=11-14 \GeV ^2$}\\
\hline
$\Delta x $ range & $\langle x\rangle$ & $\Delta\sigma / \Delta x \;
 (\rm pb )$ & $\Delta\sigma / \Delta x \;(\rm pb )$ &  $\Delta \sigma / \Delta x
\;( \rm pb )$ & $\Delta \sigma / \Delta x \;(\rm pb  )$ \\
\hline
$0.006-0.023$	&$0.013$& $103.6\pm 8.5\pm 14.5$&$ 108.2\pm10.8\pm15.4$& $106.4\pm10.2\pm10.4$&$  115.3\pm7.8\pm16.4$\\
$0.023-0.040$	&$0.031$& $\phantom{0}63.4\pm5.3\pm\phantom{0}4.3$&$\phantom{0}67.0\pm\phantom{0}6.5\pm\phantom{0}8.4$&$\phantom{0}63.9\pm\phantom{0}6.3\pm\phantom{0}5.2$&$\phantom{0}69.6\pm4.5\pm\phantom{0}7.7$\\
$0.040-0.060$	&$0.050$& $\phantom{0}52.0\pm4.0\pm\phantom{0}2.9$&$	\phantom{0}55.7\pm\phantom{0}5.1\pm\phantom{0}7.8$&$\phantom{0}53.4\pm\phantom{0}5.5\pm\phantom{0}5.3$&$\phantom{0}52.0\pm3.6\pm\phantom{0}5.4$\\
$0.060-0.090$	&$0.075$&	$\phantom{0}45.0\pm3.4\pm\phantom{0}2.2$&$	\phantom{0}47.7\pm\phantom{0}4.4\pm\phantom{0}6.1$&$\phantom{0}44.0\pm\phantom{0}4.0\pm\phantom{0}4.4$&$\phantom{0}	   43.8\pm2.8\pm\phantom{0}5.3$\\
$0.090-0.120$	&$0.10\phantom{0}$&	$\phantom{0}40.2\pm2.9\pm\phantom{0}1.8$&$	\phantom{0}40.8\pm\phantom{0}3.9\pm\phantom{0}5.3$&$	\phantom{0}39.3\pm\phantom{0}3.6\pm\phantom{0}4.3$&$\phantom{0}	   39.6\pm2.8\pm\phantom{0}5.1$\\
$0.120-0.160$	&$0.14\phantom{0}$&	$\phantom{0}37.9\pm2.8\pm\phantom{0}2.3$&$	\phantom{0}37.4\pm\phantom{0}3.4\pm\phantom{0}5.3$&$	\phantom{0}36.9\pm\phantom{0}3.4\pm\phantom{0}4.2$&$\phantom{0}	   37.9\pm2.5\pm\phantom{0}4.4$\\
$0.160-0.205$	&$0.18\phantom{0}$&	$\phantom{0}34.8\pm2.5\pm\phantom{0}1.3$&$	\phantom{0}33.9\pm\phantom{0}2.9\pm\phantom{0}4.7$&$	\phantom{0}33.0\pm\phantom{0}2.9\pm\phantom{0}4.3$&$\phantom{0}	   35.8\pm2.4\pm\phantom{0}4.3$\\
$0.205-0.260$	&$0.23\phantom{0}$&	$\phantom{0}33.3\pm2.4\pm\phantom{0}1.3$&$	\phantom{0}31.9\pm\phantom{0}3.1\pm\phantom{0}4.8$&$	\phantom{0}31.6\pm\phantom{0}2.8\pm\phantom{0}4.0$&$\phantom{0}	   32.6\pm2.0\pm\phantom{0}4.4$\\
$0.260-0.330$	&$0.29\phantom{0}$& $\phantom{0}29.8\pm 2.1\pm \phantom{0}1.7$&$\phantom{0}29.1\pm\phantom{0}2.9\pm\phantom{0}4.0$&$	\phantom{0}28.3\pm\phantom{0}2.7\pm\phantom{0}3.7$&$\phantom{0}	   30.7\pm1.9\pm\phantom{0}4.4$\\
$0.330-0.400$	&$0.36\phantom{0}$&	$\phantom{0}29.9\pm 2.1\pm\phantom{0}1.3$&	$\phantom{0}27.5\pm\phantom{0}2.7\pm\phantom{0}4.0$&$	\phantom{0}25.5\pm\phantom{0}2.6\pm\phantom{0}3.8$&$\phantom{0}	   29.5\pm1.8\pm\phantom{0}3.9$\\
					
\hline
\multicolumn{6} {|c|}{$Q^2=14-20 \GeV^2$}\\
\hline
$0.006-0.023$	&$0.013$	  &$	97.0\pm8.5\pm12.7$&$	100.4\pm10.2\pm9.8$&$	102.4\pm10.4\pm11.1$&$ 113.0\pm7.7\pm11.5$\\
$0.023-0.040$	&$0.031$          &$	59.3\pm4.9\pm\phantom{0}3.6$&$\phantom{0}	64.0\pm\phantom{0}6.5\pm4.9$&$\phantom{0}	64.2\pm\phantom{0}7.1\pm\phantom{0}6.2$&$\phantom{0}	72.2\pm5.2\pm\phantom{0}7.8$\\
$0.040-0.060$	&$0.050$	  &$	49.6\pm4.3\pm\phantom{0}5.7$&$\phantom{0}	50.8\pm\phantom{0}5.2\pm5.0$&$\phantom{0}	53.4\pm\phantom{0}5.3\pm\phantom{0}4.3$&$\phantom{0}	55.8\pm3.6\pm\phantom{0}2.7$\\
$0.060-0.090$	&$0.075$  	  &$	40.8\pm2.9\pm\phantom{0}1.7$&$\phantom{0}	42.5\pm\phantom{0}3.9\pm1.9$&$\phantom{0}	44.5\pm\phantom{0}4.2\pm\phantom{0}2.6$&$\phantom{0}	47.8\pm2.9\pm\phantom{0}3.1$\\
$0.090-0.120$	&$0.10\phantom{0}$&$	36.1\pm2.9\pm\phantom{0}1.9$&$\phantom{0}	38.0\pm\phantom{0}3.6\pm2.3$&$\phantom{0}	37.9\pm\phantom{0}3.6\pm\phantom{0}3.2$&$\phantom{0}	41.5\pm2.4\pm\phantom{0}2.0$\\
$0.120-0.160$	&$0.14\phantom{0}$&$	32.5\pm2.2\pm\phantom{0}1.5$&$\phantom{0}	35.7\pm\phantom{0}3.0\pm1.6$&$\phantom{0}	36.7\pm\phantom{0}3.3\pm\phantom{0}2.1$&$\phantom{0}	37.7\pm2.3\pm\phantom{0}1.7$\\
$0.160-0.205$	&$0.18\phantom{0}$&$	31.8\pm2.3\pm\phantom{0}1.3$&$\phantom{0}	31.9\pm\phantom{0}2.7\pm1.7$&$\phantom{0}	33.3\pm\phantom{0}3.0\pm\phantom{0}1.7$&$\phantom{0}	34.7\pm2.2\pm\phantom{0}1.5$\\
$0.205-0.260$	&$0.23\phantom{0}$&$	30.8\pm2.1\pm\phantom{0}1.0$&$\phantom{0}	30.6\pm\phantom{0}2.8\pm1.3$&$\phantom{0}	32.2\pm\phantom{0}2.8\pm\phantom{0}1.8$&$\phantom{0}	34.4\pm2.0\pm\phantom{0}1.5$\\
$0.260-0.330$	&$0.29\phantom{0}$&$	27.7\pm2.0\pm\phantom{0}1.0$&$\phantom{0}	27.6\pm\phantom{0}2.3\pm1.2$&$\phantom{0}30.0\pm\phantom{0}2.3\pm\phantom{0}1.7$&$\phantom{0}	31.2\pm1.9\pm\phantom{0}1.2$\\
$0.330-0.400$	&$0.36\phantom{0}$&$	28.1\pm2.1\pm\phantom{0}1.1$&$\phantom{0}	26.8\pm\phantom{0}2.4\pm1.3$&$\phantom{0}	29.4\pm\phantom{0}2.8\pm\phantom{0}1.3$&$\phantom{0}	29.0\pm1.9\pm\phantom{0}1.0$\\
$0.400-0.467$	&$0.42\phantom{0}$&$	26.7\pm2.1\pm\phantom{0}2.1$&$\phantom{0}	25.2\pm\phantom{0}2.5\pm1.8$&$\phantom{0}	26.8\pm\phantom{0}2.7\pm\phantom{0}1.7$&$\phantom{0}	27.8\pm1.9\pm\phantom{0}1.1$\\

\hline
\multicolumn{6} {|c|}{$Q^2=20-34 \GeV^2$}\\
\hline
$0.023-0.040$	&$0.031$  	 &$	48.6\pm5.4\pm3.6$&$	54.3\pm6.5\pm3.3$&$	37.4\pm3.6\pm2.4$&$	39.5\pm2.6\pm2.0$\\
$0.040-0.060$	&$0.050$	 &$	41.9\pm4.1\pm4.1$&$	36.0\pm3.4\pm2.4$&$	37.4\pm3.6\pm2.4$&$	39.5\pm2.6\pm2.0$\\
$0.060-0.090$	&$0.075$	 &$	35.8\pm2.8\pm1.7$&$	36.0\pm3.4\pm2.4$&$	37.4\pm3.6\pm2.4$&$	39.5\pm2.6\pm2.0$\\
$0.090-0.120$	&$0.10\phantom{0}$&$	32.8\pm2.7\pm2.2$&$	32.1\pm3.3\pm1.4$&$	32.4\pm3.3\pm1.9$&$	36.3\pm2.6\pm1.3$\\
$0.120-0.160$	&$0.14\phantom{0}$&$	29.1\pm2.5\pm1.7$&$	30.6\pm2.8\pm2.4$&$	32.1\pm3.0\pm2.5$&$	33.6\pm2.2\pm1.2$\\
$0.160-0.205$	&$0.18\phantom{0}$&$	27.7\pm2.1\pm1.1$&$	28.1\pm2.6\pm1.1$&$	28.6\pm2.6\pm1.7$&$	32.2\pm2.0\pm1.2$\\
$0.205-0.260$	&$0.23\phantom{0}$&$	25.0\pm1.7\pm1.0$&$	24.2\pm2.0\pm1.0$&$	25.0\pm2.2\pm1.3$&$	27.4\pm1.6\pm1.2$\\
$0.260-0.330$	&$0.29\phantom{0}$&$	25.0\pm1.7\pm1.0$&$	24.2\pm2.0\pm1.0$&$	25.0\pm2.2\pm1.3$&$	27.4\pm1.6\pm1.2$\\
$0.330-0.400$	&$0.36\phantom{0}$&$	24.5\pm1.8\pm1.2$&$	23.3\pm2.2\pm1.0$&$	24.3\pm2.2\pm1.4$&$	25.6\pm1.6\pm0.9$\\
$0.400-0.467$	&$0.42\phantom{0}$&$        23.6\pm1.8\pm1.5$&$	22.4\pm2.0\pm1.1$&$	23.5\pm2.4\pm1.2$&$	24.5\pm1.7\pm1.0$\\
$0.467-0.556$	&$0.49\phantom{0}$&$	22.8\pm1.6\pm2.5$&$	22.4\pm2.1\pm2.9$&$	24.0\pm2.3\pm2.3$&$	24.5\pm1.5\pm1.5$\\
\hline
\end{tabular}
\end{center}
\caption{ Cross sections $\Delta\sigma / \Delta x$ as a function of $x$ for the reaction \eeh {}
for the four average values of \rts ,
in three $ \Q2 $ intervals. The first uncertainty is statistical, the second systematic.}
\label{table2}
}
\end{table}

\hspace*{-0.5cm} 
\begin{table}{\footnotesize \begin{center} 
\begin{tabular}{|c|c|c|c|c|c|c|c|c|c|c|}
\cline{2-11}
\multicolumn{1} {c} {} & \multicolumn{10}{|c|}{$Q^2=11-14 \GeV^2$}\\

\hline
 \multicolumn{1} {|c|} {}& $0.006$&$0.023$&$0.040$&$0.060$&$0.090$&$0.120$&$0.160$&$0.205$&$0.260$&$0.330$\\
 \multicolumn{1} {|c|} {$x$ range}& $-$&$-$&$-$&$-$&$-$&$-$&$-$&$-$&$-$&$-$\\  
 \multicolumn{1} {|c|} {}& 0.023 &0.040 &0.060 &0.090 &0.120 &0.160 &0.205 &0.260 &0.330 &0.400 \\ 
\hline

$0.006-0.023$&  1.00&        &        &        &        &        &        &        &        &        \\   
\hline      
$0.023-0.040$&  0.92&    1.00&        &        &        &        &        &        &        &        \\         
\hline
$0.040-0.060$&  0.75&    0.92&    1.00&        &        &        &        &        &        &        \\        
\hline
$0.060-0.090$&  0.55&    0.79&    0.96&    1.00&        &        &        &        &        &        \\         
\hline
$0.090-0.120$&  0.45&    0.67&    0.85&    0.94&    1.00&        &        &        &        &        \\         
\hline
$0.120-0.160$&  0.39&    0.62&    0.78&    0.89&    0.98&    1.00&        &        &        &        \\         
\hline
$0.160-0.205$&  0.28&    0.50&    0.67&    0.80&    0.95&    0.97&    1.00&        &        &        \\         
\hline
$0.205-0.260$&  0.25&    0.45&    0.59&    0.69&    0.84&    0.87&    0.94&    1.00&        &        \\        
\hline
$0.260-0.330$&  0.22&    0.38&    0.49&    0.59&    0.79&    0.82&    0.91&    0.99&    1.00&        \\         
\hline
$0.330-0.400$&  0.17&    0.36&    0.50&    0.60&    0.77&    0.78&    0.86&    0.98&    0.97&    1.00\\         
\hline
\end{tabular}
 
\vspace{3ex}

\begin{tabular}{|c|c|c|c|c|c|c|c|c|c|c|c|}
\cline{2-12}
\multicolumn{1}{c}{} & \multicolumn{11}{|c|}{$Q^2=14-20 \GeV ^2$}\\

\hline
 \multicolumn{1} {|c|} {}&$ 0.006$&$0.023$&$0.040$&$0.060$&$0.090$&$0.120$&$0.160$&$0.205$&$0.260$&$0.330$&$0.400$\\ 
 \multicolumn{1} {|c|} {$x$ range}& $-$&$-$&$-$&$-$&$-$&$-$&$-$&$-$&$-$&$-$&$-$\\   
 \multicolumn{1} {|c|} {}& 0.023 &0.040 &0.060 &0.090 &0.120 &0.160 &0.205 &0.260 &0.330 &0.400 &0.467 \\ 
\hline

$0.006-0.023$&  1.00&        &        &        &        &        &        &        &        &        &         \\   
\hline
$0.023-0.040$&  0.87&    1.00&        &        &        &        &        &        &        &        &         \\   
\hline
$0.040-0.060$&  0.74&    0.95&    1.00&        &        &        &        &        &        &        &         \\   
\hline 
$0.060-0.090$&  0.55&    0.82&    0.93&    1.00&        &        &        &        &        &        &         \\   
\hline
$0.090-0.120$&  0.54&    0.79&    0.89&    0.98&    1.00&        &        &        &        &        &         \\  
\hline
$0.120-0.160$&  0.36&    0.63&    0.73&    0.89&    0.93&    1.00&        &        &        &        &         \\  
\hline
$0.160-0.205$&  0.31&    0.60&    0.71&    0.85&    0.89&    0.98&    1.00&        &        &        &         \\ 
\hline  
$0.205-0.260$&  0.24&    0.49&    0.59&    0.75&    0.81&    0.95&    0.95&    1.00&        &        &         \\ 
\hline  
$0.260-0.330$&  0.18&    0.40&    0.50&    0.64&    0.71&    0.87&    0.88&    0.97&    1.00&        &         \\ 
\hline  
$0.330-0.400$&  0.21&    0.40&    0.47&    0.56&    0.65&    0.79&    0.79&    0.91&    0.97&    1.00&         \\ 
\hline  
$0.400-0.467$&  0.11&    0.31&    0.37&    0.47&    0.55&    0.71&    0.71&    0.84&    0.91&    0.96&    1.00 \\  
\hline
\end{tabular}

\vspace{3ex}

\begin{tabular}{|c|c|c|c|c|c|c|c|c|c|c|c|}
\cline{2-12}
\multicolumn{1}{c}{} & \multicolumn{11}{|c|}{$Q^2=20-34 \GeV^2$}\\
\hline
 \multicolumn{1} {|c|} {}&$0.023$&$0.040$&$0.060$&$0.090$&$0.120$&$0.160$&$0.205$&$0.260$&$0.330$&$0.400$&$0.467$\\
 \multicolumn{1} {|c|} {$x$ range}& $-$&$-$&$-$&$-$&$-$&$-$&$-$&$-$&$-$&$-$&$-$\\    
 \multicolumn{1} {|c|} {}&0.040 &0.060 &0.090 &0.120 &0.160 &0.205 &0.260 &0.330 &0.400 &0.467 &0.556\\ 
\hline

$0.023-0.040$&  1.00&        &        &        &        &        &        &        &        &        &        \\
\hline
$0.040-0.060$&  0.92&    1.00&        &        &        &        &        &        &        &        &        \\
\hline
$0.060-0.090$&  0.87&    0.96&    1.00&        &        &        &        &        &        &        &        \\
\hline
$0.090-0.120$&  0.80&    0.88&    0.97&    1.00&        &        &        &        &        &        &        \\
\hline
$0.120-0.160$&  0.66&    0.75&    0.89&    0.96&    1.00&        &        &        &        &        &        \\
\hline
$0.160-0.205$&  0.62&    0.71&    0.82&    0.91&    0.97&    1.00&        &        &        &        &        \\
\hline
$0.205-0.260$&  0.58&    0.64&    0.76&    0.86&    0.93&    0.98&    1.00&        &        &        &        \\
\hline
$0.260-0.330$&  0.47&    0.54&    0.66&    0.77&    0.88&    0.95&    0.97&    1.00&        &        &        \\
\hline
$0.330-0.400$&  0.43&    0.49&    0.61&    0.71&    0.81&    0.88&    0.92&    0.97&    1.00&        &        \\
\hline
$0.400-0.467$&  0.38&    0.45&    0.55&    0.63&    0.73&    0.78&    0.83&    0.91&    0.98&    1.00&        \\
\hline
$0.467-0.556$&  0.39&    0.44&    0.54&    0.63&    0.73&    0.78&    0.84&    0.91&    0.98&    0.99&    1.00\\
\hline
\end{tabular}
\caption{ Correlation matrices obtained with the PYTHIA Monte Carlo
 for the data at $\rts = 189 \GeV $ for the three $ \Q2 $ intervals.}
\label{table2a}
\end{center}
 }
\end{table}

\hspace*{-0.5cm} 
\begin{table}{\footnotesize \begin{center} 

\begin{tabular}{|c|c|c|c|c|c|c|c|c|c|c|}
\cline{2-11}
\multicolumn{1} {c} {} & \multicolumn{10}{|c|}{$Q^2=11-14 \GeV ^2$}\\


\hline
 \multicolumn{1} {|c|} {}& 0.006 &0.023 &0.040 &0.060 &0.090 &0.120 &0.160 &0.205 &0.260 &0.330\\ 
 \multicolumn{1} {|c|} {$x$ range}& $-$ &$-$ &$-$ &$-$ &$-$ &$-$ &$-$ &$-$ &$-$ &$-$ \\ 
 \multicolumn{1} {|c|} {}& 0.023 &0.040 &0.060 &0.090 &0.120 &0.160 &0.205 &0.260 &0.330 &0.400\\
\hline

0.006$-$0.023&   1.00 &           &           &        &        &        &        &        &        &        \\   
\hline      
 0.023$-$0.040&   0.91&    1.00&           &        &        &        &        &        &        &        \\  
\hline      
 0.040$-$0.060&   0.79&    0.96&    1.00&    &        &        &        &        &        &               \\        
\hline      
 0.060$-$0.090&   0.63&    0.86&    0.95&    1.00&    &        &        &        &        &        \\
 \hline      
 0.090$-$0.120&   0.52&    0.74&    0.86&    0.95&    1.00&    &        &        &        &        \\
\hline      
 0.120$-$0.160&   0.46&    0.66&    0.78&    0.90&    0.98&    1.00&    &        &        &        \\         
\hline      
 0.160$-$0.205&   0.36&    0.57&    0.69&    0.82&    0.92&    0.97&    1.00&    &        &        \\
\hline      
 0.205$-$0.260&   0.27&    0.45&    0.56&    0.69&    0.80&    0.88&    0.97&    1.00&    &        \\   
\hline      
 0.260$-$0.330&   0.24&    0.42&    0.52&    0.65&    0.73&    0.83&    0.93&    0.98&    1.00&  \\   
\hline      
 0.330$-$0.400&    0.19&    0.36&    0.45&    0.58&    0.64&    0.74&    0.86&   0.93&    0.98&  1.00 \\     
\hline      
\end{tabular}
\vspace{5mm}

\begin{tabular}{|c|c|c|c|c|c|c|c|c|c|c|c|}

\cline{2-12}
\multicolumn{1} {c} {} & \multicolumn{11}{|c|}{$Q^2=14-20 \GeV ^2$}\\

\hline
 \multicolumn{1} {|c|} {}& 0.006 &0.023 &0.040 &0.060 &0.090 &0.120 &0.160 &0.205 &0.260 &0.330 &0.400\\ 
 \multicolumn{1} {|c|} {$x$ range}& $-$ &$-$ &$-$ &$-$ &$-$ &$-$ &$-$ &$-$ &$-$ &$-$ &$-$\\ 
 \multicolumn{1} {|c|} {}& 0.023 &0.040 &0.060 &0.090 &0.120 &0.160 &0.205 &0.260 &0.330 &0.400 &0.467\\
\hline

0.006$-$0.023&   1.00&    &        &        &        &        &        &        &        &        &        \\   
\hline      
0.023$-$0.040&   0.90&    1.00&    &        &        &        &        &        &        &        &        \\   
\hline      
0.040$-$0.060&   0.76&    0.95&    1.00&    &        &        &        &        &        &        &        \\   
\hline      
0.060$-$0.090&   0.62&    0.86&    0.96&    1.00&    &        &        &        &        &        &        \\   
\hline      
0.090$-$0.120&   0.50&    0.74&    0.87&    0.96&    1.00&    &        &        &        &        &        \\  
\hline      
0.120$-$0.160&   0.43&    0.66&    0.78&    0.90&    0.97&    1.00&    &        &        &        &        \\  
\hline      
0.160$-$0.205&   0.36&    0.56&    0.68&    0.81&    0.90&    0.96&    1.00&    &        &        &        \\ 
\hline      

0.205$-$0.260&   0.29&    0.45&    0.57&    0.70&    0.78&    0.87&    0.96&    1.00&    &        &        \\ 
\hline      

0.260$-$0.330&   0.23&    0.38&    0.49&    0.60&    0.68&    0.77&    0.90&    0.97&    1.00&    &        \\ 
\hline      

0.330$-$0.400&   0.17&    0.32&    0.43&    0.56&    0.63&    0.73&    0.85&    0.95&    0.98&    1.00&    \\  
\hline      
0.400$-$0.467&   0.16&    0.28&    0.38&    0.48&    0.55&    0.64&    0.77&    0.88&    0.95&    0.98&    1.00\\  
\hline
\end{tabular}
\vspace{5mm}

\begin{tabular}{|c|c|c|c|c|c|c|c|c|c|c|c|}
\cline{2-12}
\multicolumn{1} {c} {} & \multicolumn{11}{|c|}{$Q^2=20-34 \GeV ^2$ }\\

\hline
 \multicolumn{1} {|c|} {}& 0.023 &0.040 &0.060 &0.090 &0.120 &0.160 &0.205 &0.260 &0.330 &0.400 &0.467 \\ 
 \multicolumn{1} {|c|} {$x$ range}& $-$ &$-$ &$-$ &$-$ &$-$ &$-$ &$-$ &$-$ &$-$ &$-$ &$-$\\ 
 \multicolumn{1} {|c|} {}&0.040 &0.060 &0.090 &0.120 &0.160 &0.205 &0.260 &0.330 &0.400 &0.467 &0.556\\
\hline

0.023$-$0.040&   1.00&    &        &        &        &        &        &        &        &        &        \\
\hline
0.040$-$0.060&   0.92&  1.00&    &        &        &        &        &        &        &        &        \\
\hline
0.060$-$0.090&   0.85&    0.95&    1.00&    &        &        &        &        &        &        &        \\
\hline
0.090$-$0.120&   0.73&    0.87&    0.96&    1.00&    &        &        &        &        &        &        \\
\hline
0.120$-$0.160&   0.62&    0.77&    0.88&    0.96&    1.00&    &        &        &        &        &        \\
\hline
0.160$-$0.205&   0.52&    0.66&    0.78&    0.88&    0.96&    1.00&    &        &        &        &        \\
\hline
0.205$-$0.260&   0.45&    0.61&    0.71&    0.81&    0.91&    0.98&    1.00&   &        &        &        \\ 
\hline
0.260$-$0.330&   0.35&    0.50&    0.59&    0.69&    0.79&    0.90&    0.96&    1.00&    &        &        \\
\hline
0.330$-$0.400&   0.34&    0.45&    0.54&    0.62&    0.69&    0.78&    0.85&    0.93&    1.00&    &        \\
\hline
0.400$-$0.467&   0.25&    0.37&    0.46&    0.56&    0.66&    0.78&    0.87&    0.96&    0.98&    1.00&   \\
 \hline
0.467$-$0.556&   0.25&    0.36&    0.44&    0.52&    0.61&    0.72&    0.81&    0.90&    0.97&    0.98&    1.00\\
\hline
\end{tabular}
\caption{ Correlation matrices obtained with the TWOGAM Monte Carlo
 for the data at $\rts = 189 \GeV $ for the three $ \Q2 $ intervals.}
\label{table2abis}
\end{center} 
}
\end{table}

\newpage 
 
\begin{table}[h!]
\begin{center}
\begin{tabular}{|l|c|}
\hline
Source of systematic uncertainties & Uncertainty in \% \\
\hline

Tagging calorimeter polar angle & $0.9- 8.0$\\
Tagging calorimeter energy & $1.6-2.5$\\
Anti-tag energy & $0.4$\\
Number of particles & $0.2-2.6$\\
Total energy in the calorimeters& $0.2$ \\
Trigger efficiency & $1.5$\\
Monte Carlo statistics&$ 0.2-0.5$ \\
Model dependence & $0.7-9.9$\\

\hline

\end{tabular}
\end{center}
\caption{ Systematic uncertainties on the measured cross sections.}
\label{table3}
\end{table}

\newpage

\begin{table}[h!]
\begin{center}
\begin{tabular}{|c|c|c|c|c|c|c|}
\hline

$x$ range & $\langle x\rangle$ & $\cal R$ & $F_2^\gamma/\alpha$\\ 

\hline
\multicolumn{4} {|c|}{$Q^2=11-14  \GeV ^2$ }\\
\hline
$0.006-0.023$	&$0.013$ & 0.92	&0.302 $\pm$0.013 $\pm$0.026 $\pm$ 	0.029	\\
$0.023-0.040$	&$0.031$ & 0.90	&0.245 $\pm$0.011 $\pm$0.021 $\pm$	0.010	\\
$0.040-0.060$	&$0.050$ & 0.88	&0.257 $\pm$0.011 $\pm$0.023 $\pm$	0.012	\\
$0.060-0.090$	&$0.075$ & 0.90	&0.296 $\pm$0.012 $\pm$0.028 $\pm$ 	0.009	\\
$0.090-0.120$	&$0.10\phantom{0}$  & 0.89	&0.315 $\pm$0.013 $\pm$0.032 $\pm$      0.007	\\
$0.120-0.160$	&$0.14\phantom{0}$  & 0.90	&0.365 $\pm$0.015 $\pm$0.038 $\pm$ 	0.008	\\
$0.160-0.205$	&$0.18\phantom{0}$  & 0.88	&0.399 $\pm$0.017 $\pm$0.043 $\pm$ 	0.007	\\
$0.205-0.260$	&$0.23\phantom{0}$  & 0.89	&0.441 $\pm$0.018 $\pm$0.049 $\pm$ 	0.011	\\
$0.260-0.330$	&$0.29\phantom{0}$  & 0.88	&0.483 $\pm$0.020 $\pm$0.054 $\pm$	0.013	\\
$0.330-0.400$	&$0.36\phantom{0}$  & 0.89	&0.536 $\pm$0.023 $\pm$0.059 $\pm$	0.025	\\

\hline
\multicolumn{4} {|c|}{$Q^2=14-20 \GeV ^2$}\\
\hline
$0.006-0.023$	&$0.013$	    &0.93 &0.310 $\pm$0.014 $\pm$0.020 $\pm$ 	0.028	\\
$0.023-0.040$	&$0.031$          	&0.88 &0.258 $\pm$0.012 $\pm$0.014 $\pm$	0.018	\\
$0.040-0.060$	&$0.050$	       &0.90 &0.275 $\pm$0.012 $\pm$0.014 $\pm$ 	0.017	\\
$0.060-0.090$	&$0.075$  	       &0.90 &0.288 $\pm$0.011 $\pm$0.014 $\pm$	0.008	\\
$0.090-0.120$	&$0.10\phantom{0}$	&0.89 &0.316 $\pm$0.013 $\pm$0.015 $\pm$       0.014	\\
$0.120-0.160$	&$0.14\phantom{0}$	&0.90 &0.337 $\pm$0.013 $\pm$0.016 $\pm$ 	0.007	\\
$0.160-0.205$	&$0.18\phantom{0}$	&0.89 &0.381 $\pm$0.015 $\pm$0.018 $\pm$ 	0.006	\\
$0.205-0.260$	&$0.23\phantom{0}$	&0.88 &0.424 $\pm$0.017 $\pm$0.019 $\pm$ 	0.008	\\
$0.260-0.330$	&$0.29\phantom{0}$	&0.88 &0.471 $\pm$0.018 $\pm$0.020 $\pm$ 	0.009	\\
$0.330-0.400$	&$0.36\phantom{0}$	&0.87 &0.510 $\pm$0.021 $\pm$0.022 $\pm$ 	0.007	\\
$0.400-0.467$ 	&$0.42\phantom{0}$	&0.88 &0.551 $\pm$0.024 $\pm$0.024 $\pm$ 	0.026	\\

\hline
\multicolumn{4} {|c|}{$Q^2=20-34 \GeV ^2$}\\
\hline
$0.023-0.040$	&$0.031$  	  & 0.89 &0.317 $\pm$0.017 $\pm$ 0.016	$\pm$0.017	\\
$0.040-0.060$	&$0.050$	  & 0.89 &0.293 $\pm$0.015 $\pm$ 0.014	$\pm$0.010	\\
$0.060-0.090$	&$0.075$	  & 0.89 &0.314 $\pm$0.013 $\pm$ 0.015	$\pm$0.012	\\
$0.090-0.120$	&$0.10\phantom{0}$	& 0.88 &0.338 $\pm$0.016 $\pm$ 0.016	$\pm$0.018	\\
$0.120-0.160$	&$0.14\phantom{0}$	& 0.88 &0.384 $\pm$0.017 $\pm$ 0.018	$\pm$0.017	\\
$0.160-0.205$	&$0.18\phantom{0}$	& 0.88 &0.404 $\pm$0.017 $\pm$ 0.018	$\pm$0.004	\\
$0.205-0.260$	&$0.23\phantom{0}$	& 0.88 &0.446 $\pm$0.018 $\pm$ 0.020	$\pm$0.009	\\
$0.260-0.330$	&$0.29\phantom{0}$	& 0.87 &0.488 $\pm$0.019 $\pm$ 0.022	$\pm$0.006	\\
$0.330-0.400$	&$0.36\phantom{0}$	& 0.87 &0.557 $\pm$0.023 $\pm$ 0.025	$\pm$0.010	\\
$0.400-0.467$	&$0.42\phantom{0}$	& 0.87 &0.611 $\pm$0.027 $\pm$ 0.029	$\pm$0.015	\\
$0.467-0.556$ 	&$0.49\phantom{0}$	& 0.87 &0.683 $\pm$0.028 $\pm$ 0.030	$\pm$0.062	\\

\hline
\end{tabular}
\end{center}
\caption{Measured values of $F_2^\gamma/\alpha$ and the applied radiative correction factors, $\cal R $,  
in bins of $x$ for the three $Q^2$ ranges. 
The first uncertainty is statistic, the second systematic and the third is due to model dependence.}
\label{table4}
\end{table}

\begin{table}[h!]
\begin{center}
\begin{tabular}{|c|c|c|c|}
\hline
\rule{0pt}{12pt}
$Q^2$ range & $\langle Q^2\rangle$ & $\cal R $ & $ F_2^\gamma/\alpha $ \\ 

\hline
\multicolumn{4} {|c|}{$x=0.01-0.1$}\\
\hline
$11-14$	&12.4 &0.89 &0.278 $\pm$ 0.006 $\pm$ 0.028 $\pm$ 	0.013	\\
$14-20$	&16.7 &0.89 &0.287 $\pm$ 0.006 $\pm$ 0.015 $\pm$       0.015	\\
$20-34$	&25.5 &0.88 &0.316 $\pm$ 0.008 $\pm$ 0.016 $\pm$ 	0.013	\\

\hline
\multicolumn{4} {|c|}{$x=0.1-0.2$}\\
\hline
$11-14$	&12.4 &0.88 &0.377 $\pm$ 0.010 $\pm$ 0.039 $\pm$ 	0.008	\\
$14-20$	&16.7 &0.88 &0.355 $\pm$ 0.009 $\pm$ 0.017 $\pm$       0.005	\\
$20-34$	&25.5 &0.88 &0.399 $\pm$ 0.011 $\pm$ 0.019 $\pm$ 	0.010	\\

\hline
\multicolumn{4} {|c|}{$x=0.2-0.3$}\\
\hline
$11-14$	&12.4	&0.88 &0.464  $\pm$ 0.015 $\pm$ 0.051 $\pm$ 	0.009	\\
$14-20$	&16.7	&0.88 &0.442  $\pm$ 0.013 $\pm$ 0.020 $\pm$    0.003	\\
$20-34$	&25.5	&0.87 &0.477  $\pm$ 0.015 $\pm$ 0.023 $\pm$ 	0.013	\\

\hline
\multicolumn{4} {|c|}{$x=0.3-0.5$}\\
\hline
11-14	&12.4	&0.89 &0.544  $\pm$ 0.017 $\pm$ 0.061 $\pm$ 	0.019	\\
14-20	&16.7	&0.87 &0.545  $\pm$ 0.014 $\pm$ 0.024 $\pm$    0.012	\\
20-34	&25.5	&0.87 &0.594  $\pm$ 0.015 $\pm$ 0.029 $\pm$ 	0.022	\\

\hline
\end{tabular}
\end{center}
\caption{The values of  $F_2^\gamma / \alpha $ in bins of $\Q2 $ for four  $x$ ranges together
with the radiative correction factor. 
The first uncertainty is statistic, the second systematic and the third is due to model dependence.}
\label{table5}
\end{table}

\newpage 

\begin{sidewaysfigure}
\begin{center}
\includegraphics[width=0.45\textwidth]{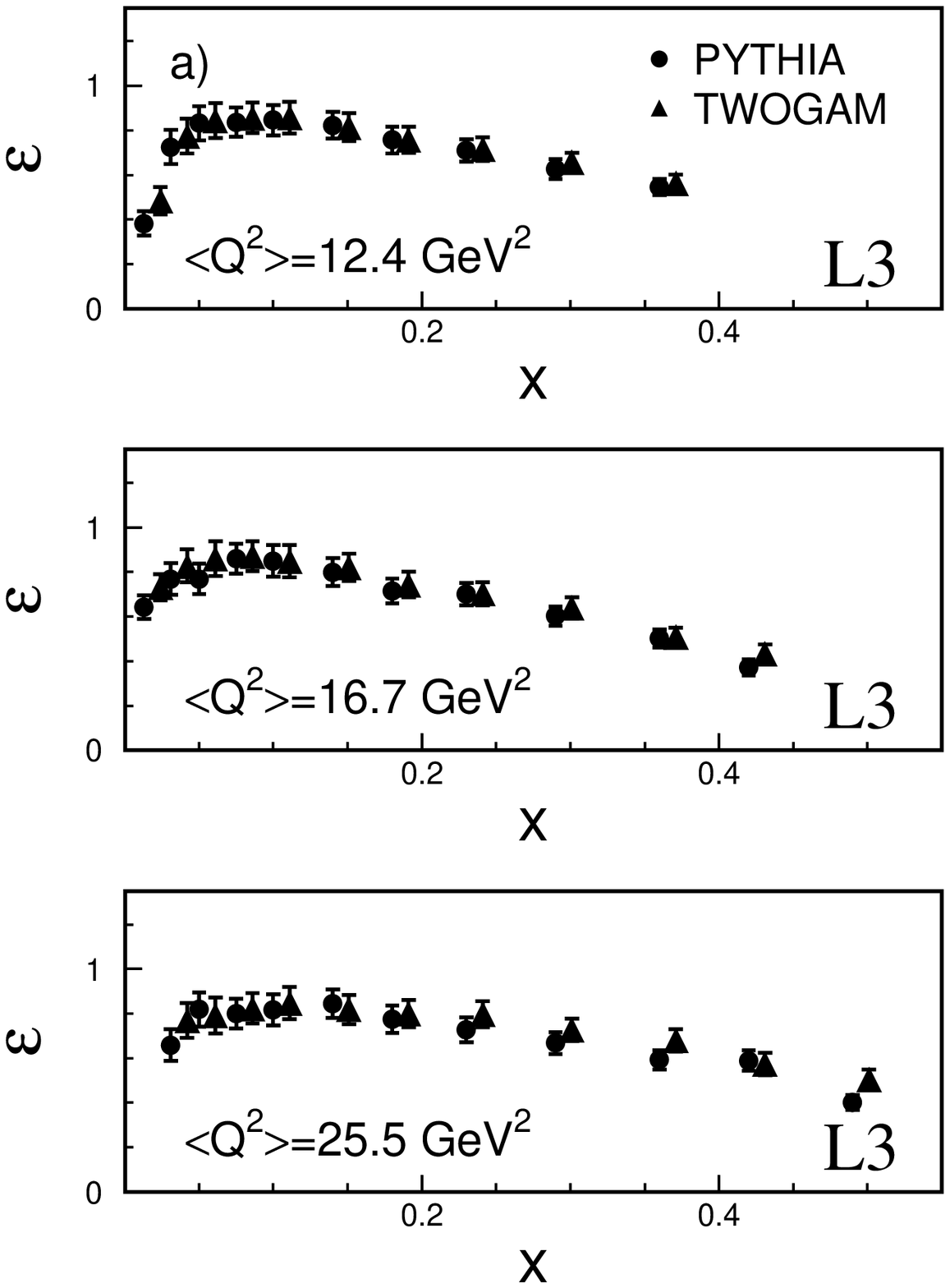}
\includegraphics[width=0.45\textwidth]{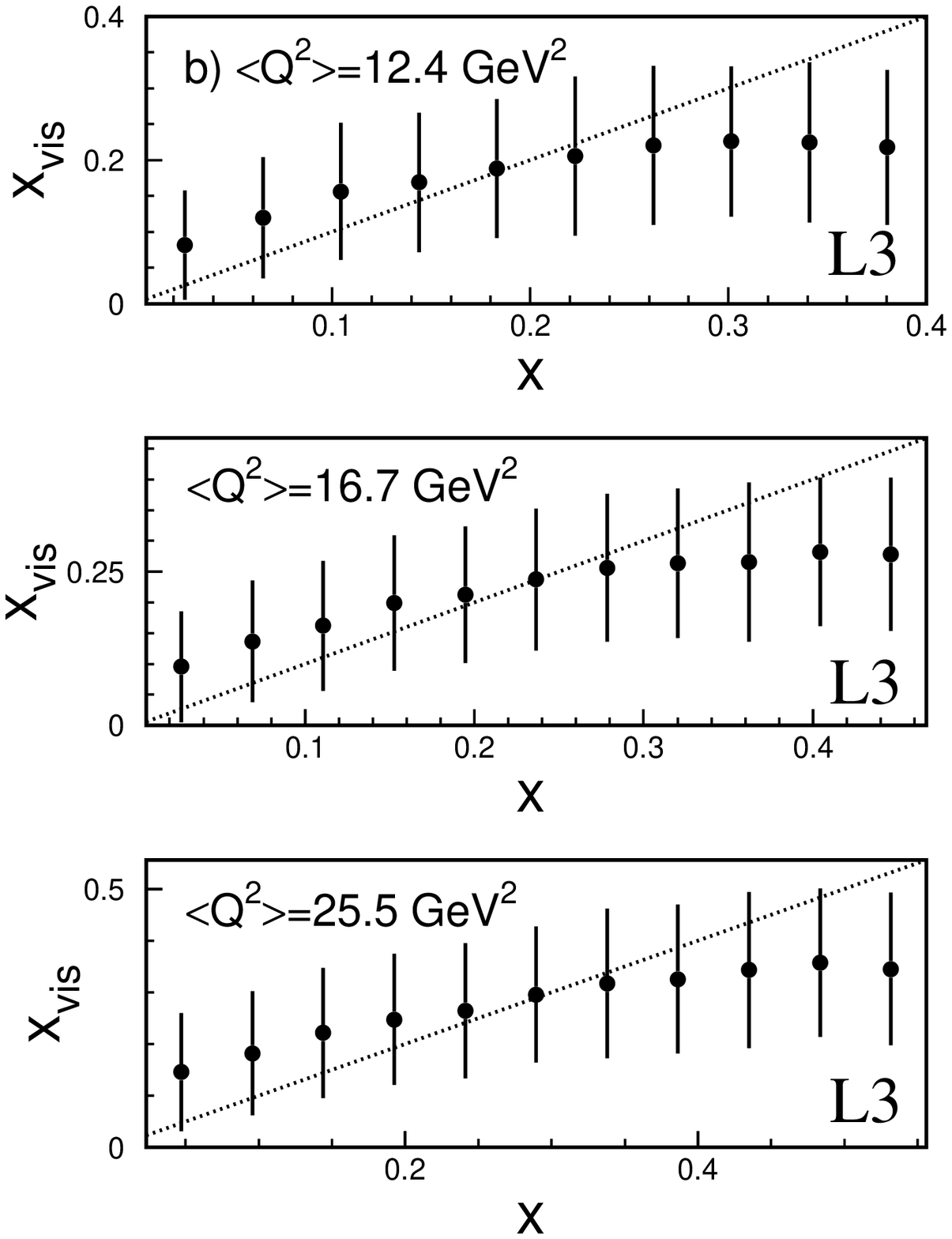}

\caption{a) The detector acceptance and selection efficiency,
$\varepsilon $, obtained by the PYTHIA and TWOGAM generators.  b)
Comparison of the reconstructed and generated value of $x$ for the
PYTHIA Monte Carlo at $\rts = 189 \GeV $ for different values of $\Q2
$. The mean observed value and the standard deviation of
$x_\mathrm{vis}$ are plotted for events generated in a given
$x$ bin.  }
\end{center} 
\label{fig4}
\end{sidewaysfigure}

\begin{figure}[htbp]
\begin{center}
\includegraphics[width=0.45\textwidth]{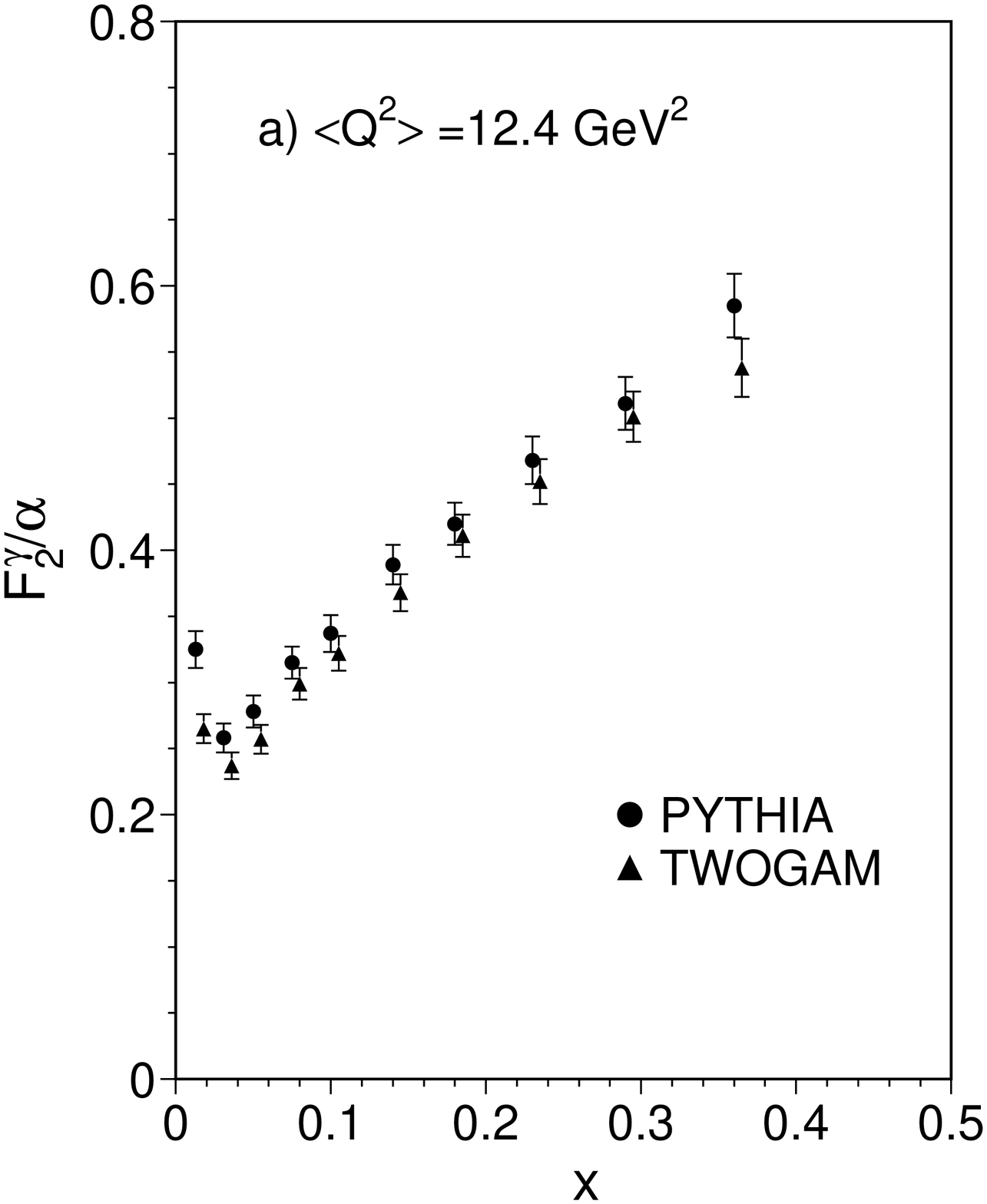}
\includegraphics[width=0.45\textwidth]{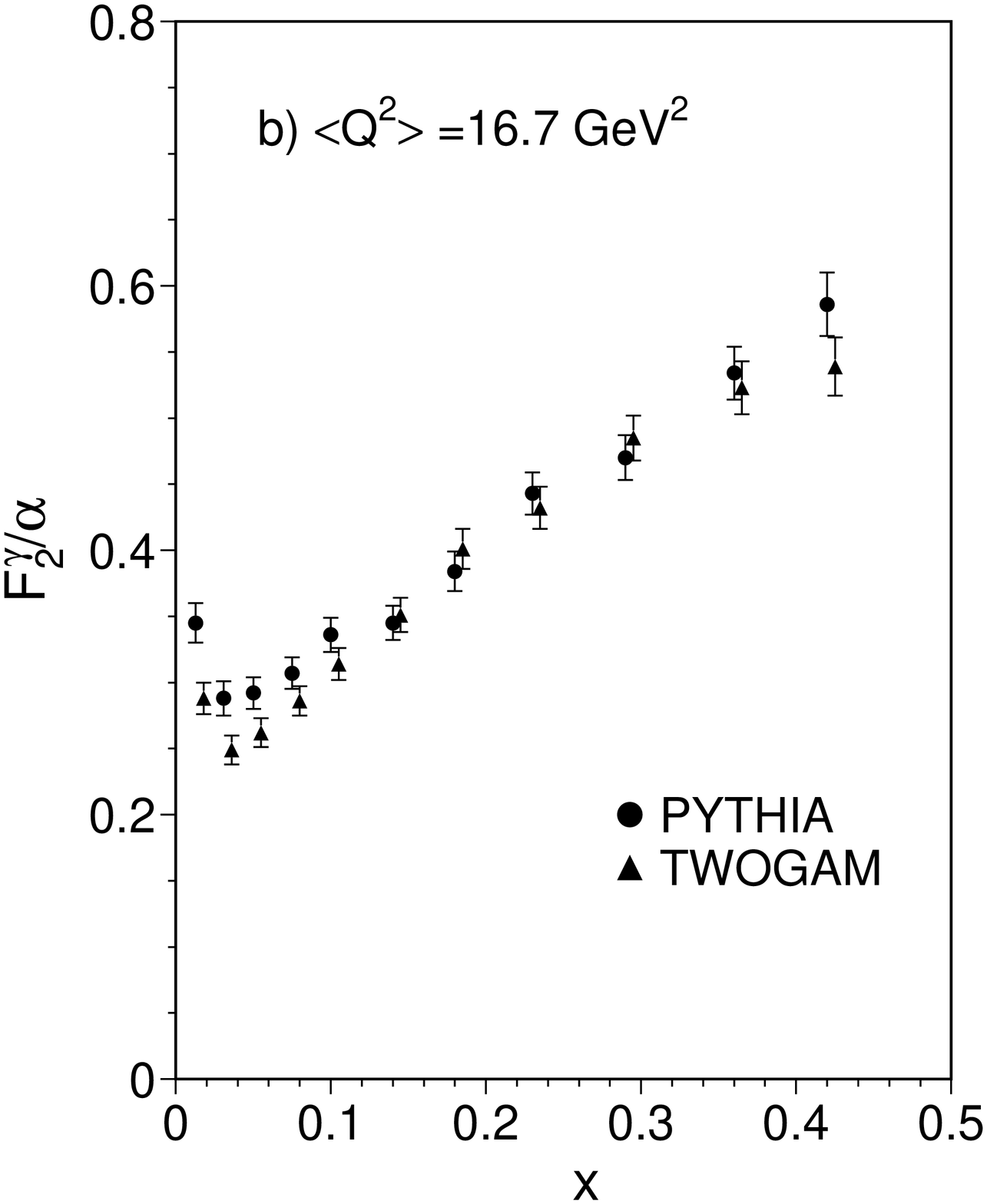}
\includegraphics[width=0.45\textwidth]{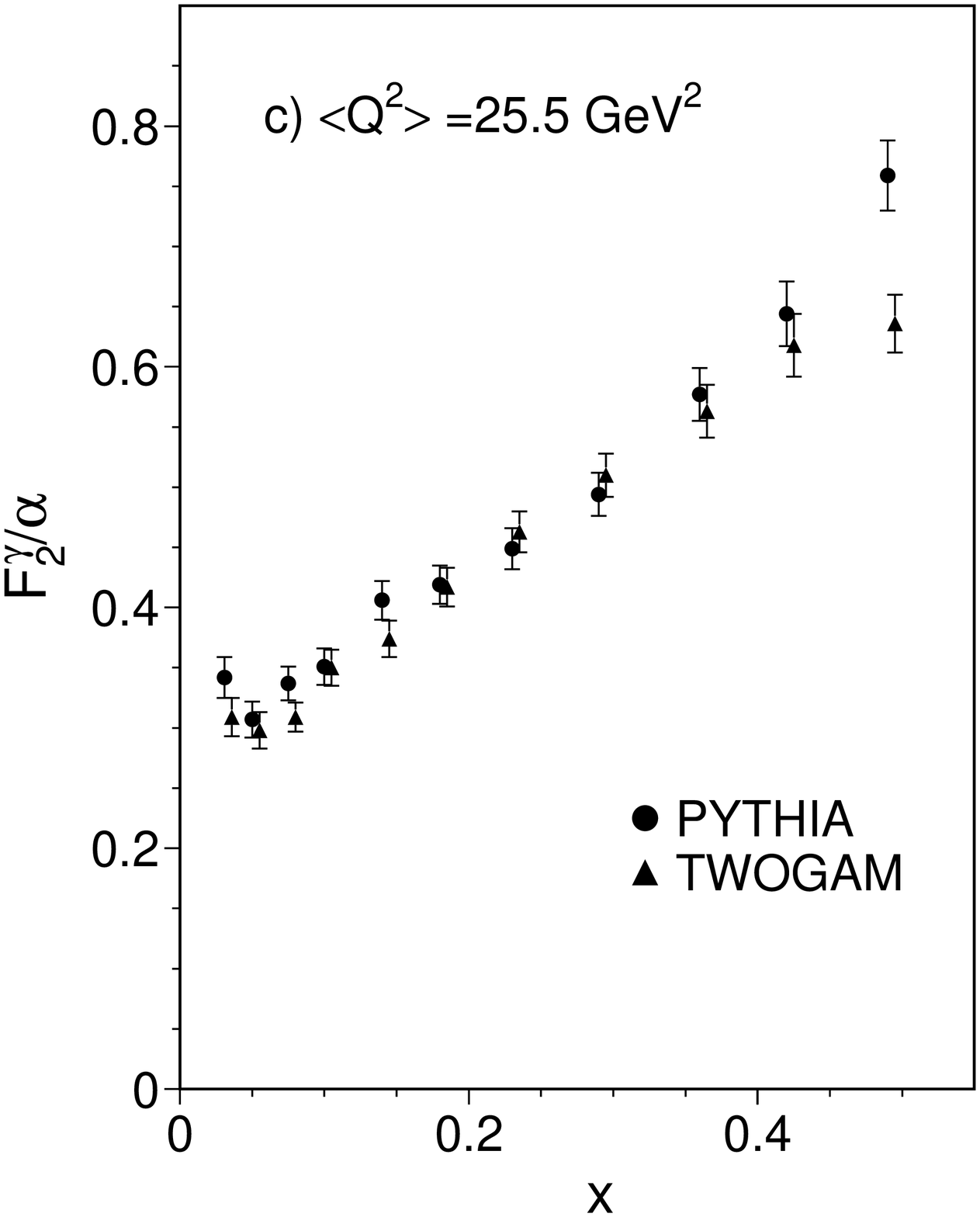}

\caption{The photon structure function $F_2^\gamma / \alpha$ obtained
with PYTHIA and TWOGAM. Only the statistical uncertainties are shown. 
For clarity, the symbols
corresponding to the two Monte Carlo generators are slightly
offset. The measurements are correlated as indicated in
Tables~\protect{\ref{table2a}} and~\protect{\ref{table2abis}}}
\end{center}
\label{fig5}
\end{figure}

\begin{figure}[htbp]
\begin{center}
\includegraphics[height=0.4\textheight]{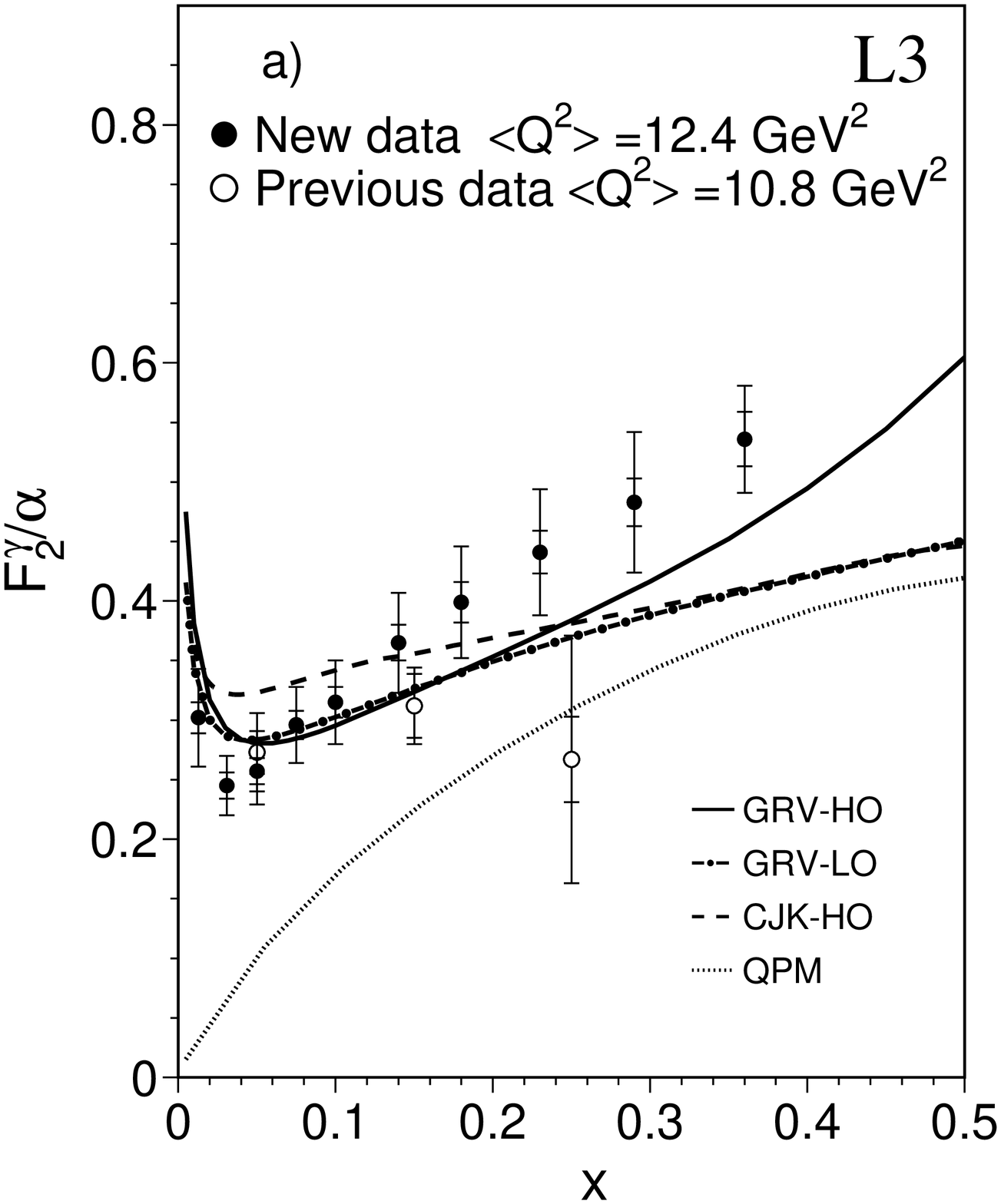}
\includegraphics[height=0.4\textheight]{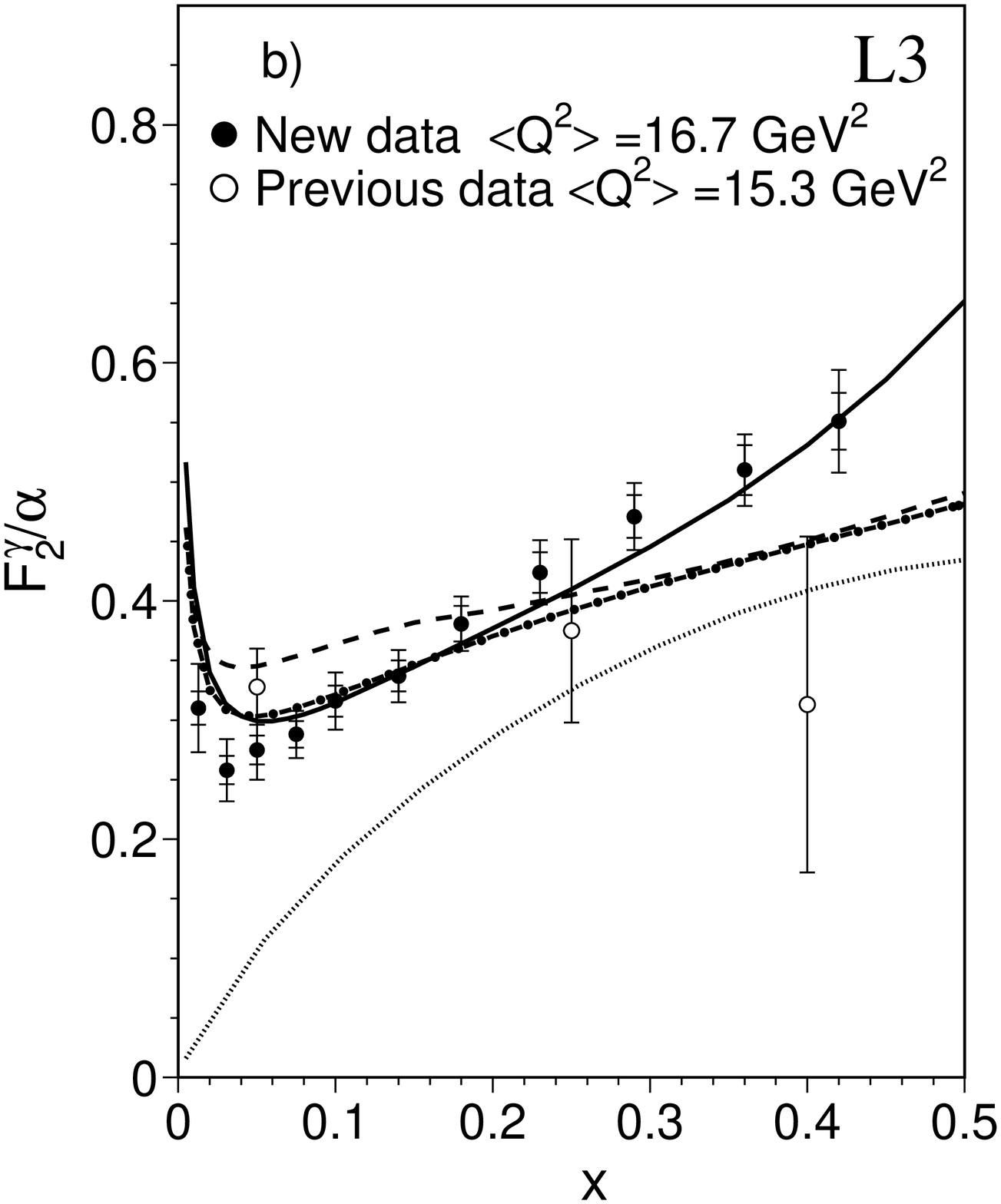}
\includegraphics[height=0.4\textheight]{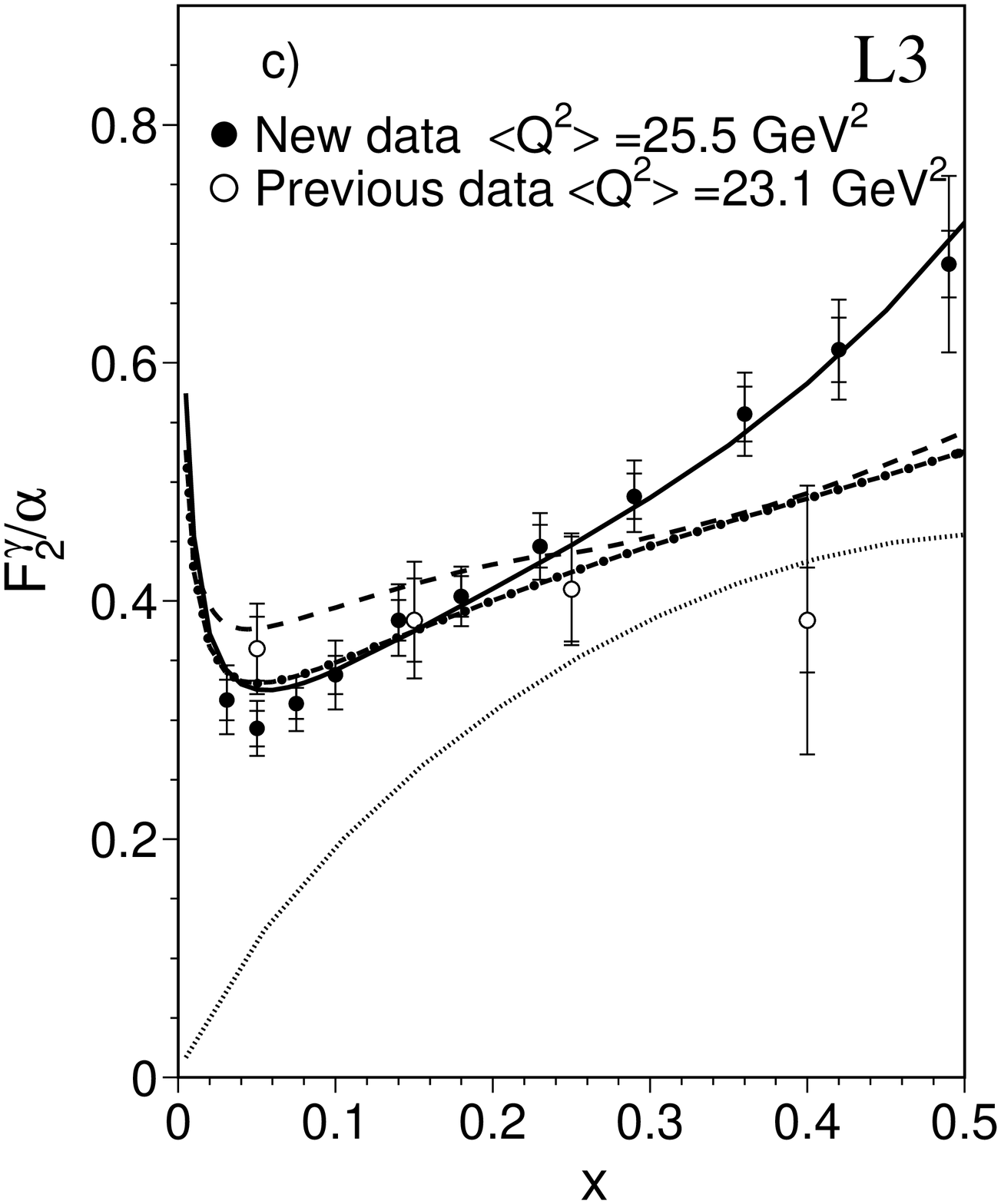}

\end{center}
\caption{ The photon structure function $F_2^\gamma / \alpha$ as a
function of $x$ for the three $\Q2 $ intervals, with statistical and
systematic uncertainties. The former are indicated by the inner error
bars. The new data are presented together with the previous results at
$\rts = 183 \GeV $\protect\cite{previous}. The predictions of the
higher-order parton density functions GRV and CJK are shown as well
as the leading-order predictions of the GRV. The changes in slope
of the CJK predictions are due to the c-quark threshold. The QPM predictions
for $\gamma \gamma \to \qq $ are also shown.}
\label{fig6}
\end{figure}

 \begin{figure}[htbp]
\begin{center}
\includegraphics[width=0.45\textwidth]{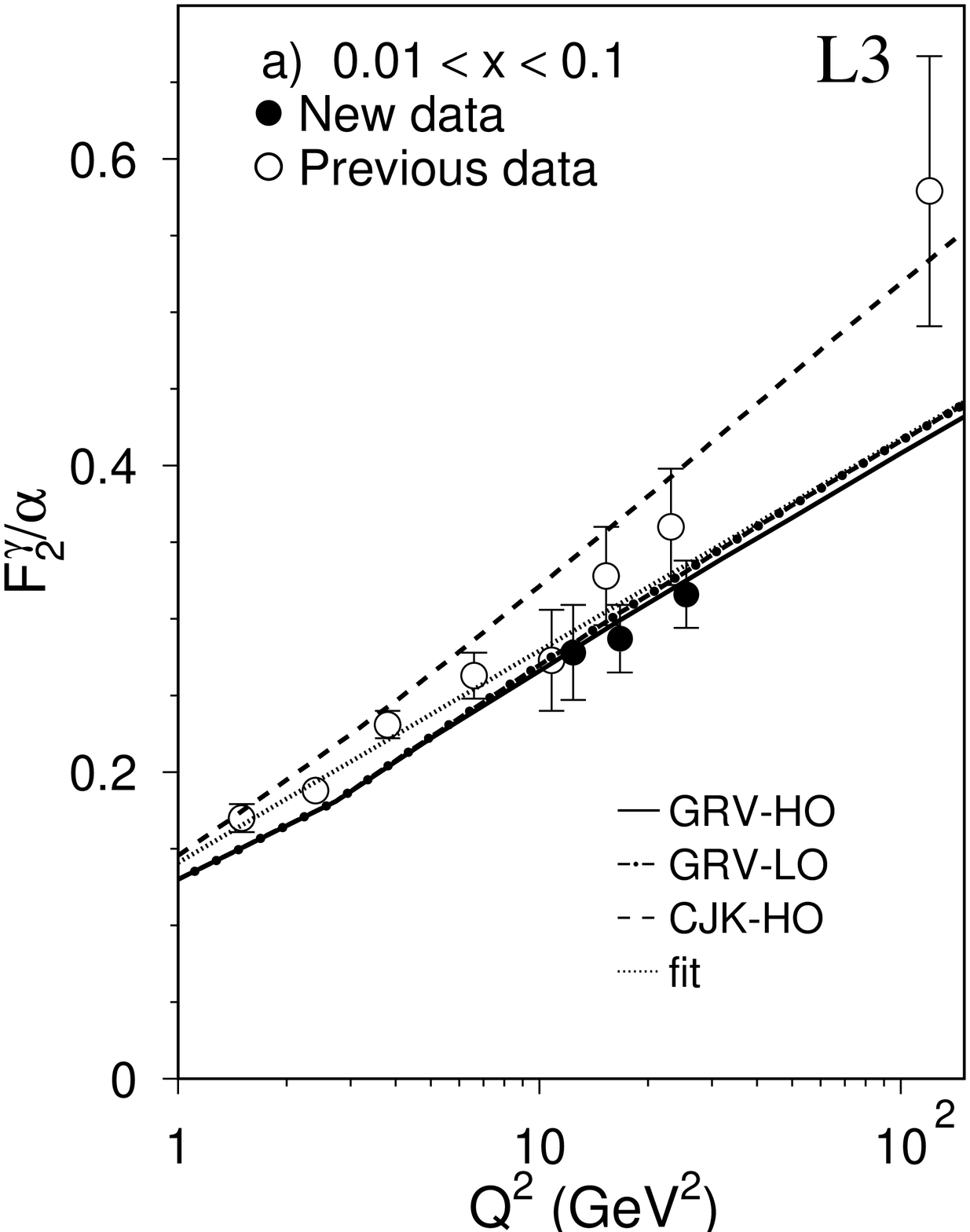}
\includegraphics[width=0.45\textwidth]{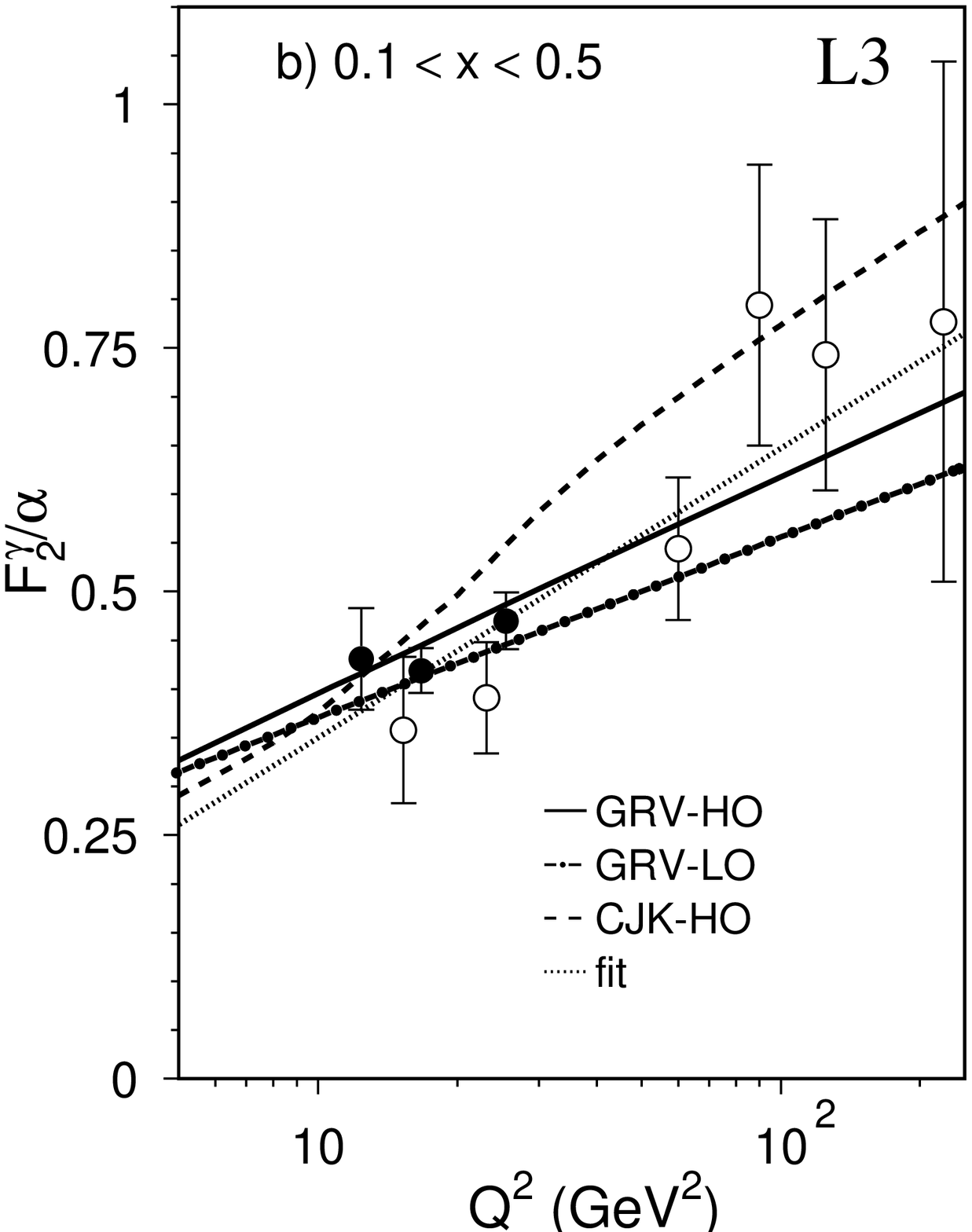}
\end{center}
\caption{Evolution of the photon structure function $F_2^\gamma /
\alpha$ as a function of $ \Q2 $ for two $x$ intervals. The results of
a fit to the data of the function $a + b (\ln{Q^2}/\GeV^2)$ are shown
together with the predictions of the higher-order parton density
functions GRV and CJK as well as the leading-order predictions of
GRV.}
\label{fig7}
\end{figure}

\end{document}